\documentclass[10pt,floatfix, prl, nopacs, twocolumn,showkeys, preprintnumbers, nofootinbib, superscriptaddress, nolongbibliography]{revtex4-2}
\usepackage[utf8]{inputenc}
\usepackage[sort&compress]{natbib}
\usepackage{ulem}
\usepackage{overpic}
\usepackage{bm}
\usepackage{times}
\usepackage{amssymb,amsbsy,amsmath,amsfonts}
\usepackage{graphicx}
\usepackage{float}
\usepackage{lipsum}
\usepackage{color}
\usepackage{booktabs}
\usepackage{makecell}
\usepackage{rotating}
\usepackage{srcltx}
\usepackage{slashed}
\usepackage{subfigure}
\usepackage{multirow}
\usepackage{verbatim}
\usepackage{hyperref}
\hypersetup{colorlinks,linkcolor=red,citecolor=red}
\usepackage{tabularx}
\usepackage{newtxtext, newtxmath}
\usepackage{physics}
\usepackage{tikz,import}
\usepackage{tikz-cd}
\usepackage{pgfplots}
\usepackage{CJKutf8}
\pgfplotsset{compat=1.14}

\newcommand{\pd}[2]{\partial_{#2}{#1}}
\newcommand{\covpd}[2]{\mathcal{D}_{#2}{#1}}
\newcommand{\anticom}[2]{\qty{#1,#2}}
\newcommand{\com}[2]{\qty[#1,#2]}

\newcommand{\uv}{\affiliation{Departamento de F\'{\i}sica Te\'orica and IFIC,
Centro Mixto Universidad de Valencia-CSIC, Parc Científic UV, C/ Catedrático José Beltrán, 2, 46980 Paterna, Spain} }

\begin{document}
\title{Revealing the $D_0^*(2300)$ two-pole structure from lattice data and the SU(3) limit}
\author{Zejian~Zhuang~(\begin{CJK*}{UTF8}{gbsn}庄泽坚\end{CJK*})}
\email[Corresponding author:~]{zejian.zhuang@uv.es}
\uv
\author{Fernando~Gil~Dom\'{\i}nguez}
\affiliation{School of Physics, University of Electronic Science and Technology of China, Chengdu 611731, China}
%
%
\author{Raquel~Molina}
\email[Corresponding author:~]{raquel.molina@ific.uv.es}
\uv
\begin{abstract}
We perform an analysis of LQCD light - charmed (pseudoscalar) meson scattering data with UChPT for pion masses ranging from $m_\pi\simeq 230$~MeV till the SU(3) limit, $m_\pi\simeq 700$~MeV. We find two poles in the non-strange isospin $I=1/2$ sector that can be related to the experimental $D_0(2300)$ resonance. At the physical pion mass, the poles are located at $\sqrt{s_0}=2094(7)(1)-i111(7)(13)$ MeV, and $2463(60)(30)-i108(14)(12)$~MeV. While the first pole, named here $D_0^*(2100)$, is always a resonance in $D\pi$ within the $1\sigma$ region, the second pole can be a resonance or virtual state close to the $D\eta, D_s\bar{K}$ thresholds. For the first time, the pion mass dependence on different chiral trajectories including SU(3) LQCD data are investigated for these poles. We find that in the $m_s=m_{s,\mathrm{phy}}$ trajectory, the $D_0^*(2100)$ resonance pole behaves similarly as the $\sigma$ resonance in $\pi\pi$ scattering, splitting into two poles, connected to the $\bar{\mathbf{3}}$ representation. Moreover, we found that the higher pole related to the experimental $D_0^*(2300)$ can be related to the $\mathbf{6}$ representation. We highlight that since this pole couples strongly to channels with hidden strangeness, its mass is fairly constant in the $\mathrm{Tr}[M]=C$ trajectory, what can be tested in future LQCD simulations. The compositeness of the $D_0^*(2100)$ state at the SU(3) limit is evaluated. Finally, other sectors are also discussed. 
\end{abstract}
\maketitle
\noindent{\itshape\bfseries Introduction.---}%
Despite several years of the discovery of the scalar and axial charmed excited states, the $D_0^*(2300)$ and the $D_1^*(2430)$~\cite{Belle:2003nsh,ParticleDataGroup:2024cfk}, the interpretations of those remains elusive. The observation of their partners in the charm-strange sector, the $D_{s0}(2317)$ and $D_{s1}(2460)$~\cite{BaBar:2003oey,CLEO:2003ggt}, near the $DK$ and $D^*K $ thresholds, poses some puzzles for the quark model expectations. In particular, these states should be $100$ MeV heavier than their charmed counterparts, while the masses quoted in the PDG are $m_{D_0^*}=2343\pm 10$ MeV, $\Gamma_{D_0^*}=229\pm 16$ MeV, $m_{D_1^*}=2412\pm9$ MeV, $\Gamma_{D_1^*}=314\pm 29$ MeV\footnote{There is also a narrow $1^+$~\cite{LHCb:2019juy} state with mass $2420$ MeV and width close to $30$ MeV in the PDG~\cite{ParticleDataGroup:2024cfk}.}. The $D_0^*(2300)$ is expected to have a mass around $2400$ MeV in the quark model~\cite{Godfrey:1985xj,Godfrey:1986wj}. According to Heavy Quark Symmetry (HQS), in the limit $m_Q\to \infty$, the properties of the $Q\bar{q}$ mesons become independent of the spin of the heavy quark and one can use the Heavy Quark Spin Symmetry (HQSS) basis, to classify the states~\cite{Isgur:1989vq,Manohar:2000dt}. Thus, one finds two doublets having $J=j_q\pm \frac{1}{2}$, with $j_q$ the full spin of the light degrees of freedom, corresponding to $\lbrace 0^+,1^+\rbrace$, and $\lbrace 1^+,2^+\rbrace$, respectively. While in the first doublet the states are expected to be broad, the ones in the second doublet should be narrow $1P$ excitations~\cite{Godfrey:1985xj,Godfrey:1986wj}. However, in the charm-strange sector, the observed $D_{s0}(2317)$ and $D_{s1}(2460)$ are found to be very narrow states~\cite{ParticleDataGroup:2024cfk}. Since the masses of their partners in the charm sector,  $D^*_0(2300)$ and the $D^*_1(2430)$, are similar, this leads to some difficulties to classify these hadrons in the spectrum. Moreover, there are strong evidences that the  $D_{s0}(2317)$ and $D_{s1}(2460)$ are predominantly molecular states, as concluded by recent precise global analyses of LQCD data within unitarized chiral-based EFT approaches~\cite{Gil-Dominguez:2023puj}, where the pole parameters from the simulation were shown to be consistent with the experimental ones after a proper extrapolation to the physical point that led to approximately $70$ \% composition of two mesons, consistently with other analysis~\cite{MartinezTorres:2014kpc,Liu:2012zya,Albaladejo:2018mhb}, and with the molecular hypothesis~\cite{Barnes:2003dj,Kolomeitsev:2003ac,Faessler:2007gv,Gamermann:2006nm,Gamermann:2007fi}. Note also that the LQCD spectrum related to the scalar and axial charm mesons can only be obtained reliably when $D^{(*)}K$ interpolators are accounted for in  simulations~\cite{Bali:2017pdv,Mohler:2013rwa,Lang:2014yfa,Liu:2012zya,Cheung:2020mql}. Naturally, one would expect that their charmed counterparts belonging to the same SU(3) multiplet have also some important $D\pi$ and $D^*\pi$ components. Hence, being plausible the two-meson molecular interpretation for the $D^*_0(2300)$ and $D_1^*(2430)$ states~\cite{Gamermann:2007fi,Du:2017zvv,Khemchandani:2023xup}. 

More intriguingly, Unitarized Chiral Perturbation Theory (UChPT) for heavy mesons predicts two $D_0^*$ poles for the $D^*_0(2300)$ state, that would arise from the $D\pi, D\eta, D_s\bar{K}$ coupled-channel interaction~\cite{Kolomeitsev:2003ac,Gamermann:2006nm,Albaladejo:2016lbb,Du:2017zvv}. Nevertheless, the higher energy pole has not yet been confirmed by LQCD for pion masses in the $200-400$ MeV range~\cite{Gayer:2021xzv,Moir:2016srx}, although there are some evidences from a recent simulation done in the SU(3) limit~\cite{Yeo:2024chk}. In~\cite{Asokan:2022usm} it is suggested that the pole is not found because it was in a hidden Riemann Sheet (R.S.). Still, the work of~\cite{Asokan:2022usm} concludes that it is difficult to determine the position of the second pole due to the fact that this is highly dependent on the parametrization used to fit the data. It also remains unknown whether the second pole found near the SU(3) limit~\cite{Yeo:2024chk} is connected to the second pole found in UChPT around the physical pion mass. 

The first LQCD simulation for $D\pi$ scattering was done in~\cite{Mohler:2012na} for small boxes. See also \cite{Moir:2016srx,Gayer:2021xzv,Yeo:2024chk,Yan:2024yuq} for subsequent refined simulations. In~\cite{Liu:2012zya} the scattering lengths of the light pseudoscalar with charmed mesons were evaluated in LQCD for a pion mass range of $300-600$ MeV, concretely, the $I=3/2$ $D\pi, D_s\pi, D_sK$, and $I=0,1$ $D\bar{K}$ channels are investigated, to predict subsequently the $I=1/2$ $D\pi$ and $I=0$ $DK$ scattering lengths from an analysis with UChPT. Their quark mass dependence is then extracted. With the LECs in UChPT fixed from this analysis, in~\cite{Albaladejo:2016lbb}, the energy levels of the~\cite{Moir:2016srx} simulation for $m_\pi=391$ MeV are successfully reproduced, providing strong evidence for the two-$D_0^*$ pole prediction in UChPT. While the lower pole is connected to the $\mathbf{\bar{3}}$ representation, the one at higher energies is related to the $\mathbf{6}$ representation \cite{Albaladejo:2016lbb}. See also the recent analysis of the LQCD data~\cite{Yan:2024yuq}  for $D\pi$ scattering~\cite{Luo:2026kui}. Moreover, in~\cite{Du:2020pui} it is shown that the scattering amplitudes obtained in~\cite{Liu:2012zya} are also consistent with recent experimental LHCb data on $B^-\to D^+\pi^-\pi^-$~\cite{LHCb:2016lxy}, reinforcing the two-$D^*_0$ pole hypothesis and showing that these experimental data are in conflict with the pole parameters quoted in the PDG~\cite{ParticleDataGroup:2024cfk}. See also~\cite{Du:2017zvv,Du:2019oki}. Furthermore, the quark model expectations for the scalar charmed meson are also incompatible with experimental data on semileptonic decays~\cite{Bigi:2007qp,Blossier:2009vy}. 

The tetraquark model has also made predictions for the excited scalar and axial charmed mesons~\cite{Cheng:2003kg,Maiani:2004vq,Maiani:2024quj}. There is here a crucial difference with the molecular model. While in the latter, the interaction in the $\mathbf{\bar{3}}$ is more attractive than that of the $\mathbf{6}$ representation, the $\mathbf{\overline{15}}$ sector is repulsive and, then, cannot generate states. This is actually in qualitative agreement with the results from the LQCD simulation of~\cite{Yeo:2024chk}. On the contrary, the interaction in the tetraquark model for the $\mathbf{\overline{15}}$ also gives rise to states when one takes into account spin one and zero diquarks to generate scalar and axial charmed mesons, being those of similar mass than the ones appearing in the $\mathbf{\bar{3}}$, and heavier than the states generated in the $\mathbf{6}$ representation~\cite{Barabanov:2020jvn,Guo:2025imr}. This phenomena has been demonstrated to be incompatible with LQCD~\cite{Gregory:2025ium}. 

Despite the remarkable success of UChPT for heavy-light meson interaction describing data from LQCD and experiment, the task of extracting precisely the quark mass dependence of the two-pole structure for the $D_0^*(2300)$ towards the SU(3) limit, considering the several sources of it provided by LQCD, i.e., two-meson scattering, charmed and light meson masses, and coupling between the light and heavy meson, dominated by $f_\pi$, remains to be done. Note that in previous studies~\cite{Gamermann:2006nm,Liu:2012zya,Albaladejo:2016lbb}, only qualitative formulas for the meson masses depending on a parameter $x$ connecting the physical point and the SU(3) limit, or chiral-based leading order (LO) formulas for the meson masses and fixed $f_\pi$, were employed. SU(3) LQCD data were not included either. In this work, we analyze the data of \cite{Moir:2016srx,Gayer:2021xzv,Yeo:2024chk} and, as novelty, make predictions on several chiral trajectories towards the SU(3) limit, including the symmetric line, for the two-pole $D^*_0(2300)$ states. Since the SU(3) limit can be reached in some chiral trajectories, as the one fixing the strange quark mass to the physical point, at as high as $m_\pi\simeq 700$ MeV, it is necessary here to go beyond the NLO ChPT formulas for the pseudoscalar light meson masses and decay constants, that are only valid till $m_\pi=450$ MeV~\cite{Molina:2020qpw}. We rely here on a recent NNLO analysis which extends~\cite{Molina:2020qpw} and has been shown to be reliably applicable till $m_\pi\simeq 700$ MeV. This article is structured as follows. First, we briefly summarize the formalism, then, we present the result for the LQCD data analysis and discuss the SU(3) limit. Finally, some conclusions are provided.

\noindent{\itshape\bfseries Formalism.---}%
The NLO interaction of light-pseudoscalars with heavy mesons is derived in UChPT in~\cite{Guo:2008gp,Meng:2022ozq,Guo:2009ct,Yao:2015qia}. The interaction depends on the pion pseudoscalar decay constant $f_\pi$, which is taken from the SU(3) LQCD data analysis up to NNLO in ChPT performed in~\cite{deconsnnlo}. This interaction also has as free-parameters the LECs $h_i$, $i=0,5$, where $h_0,h_1$ are fixed by the heavy meson masses and their splittings. The latter are taken from the one-loop NLO analysis of LQCD data with HHChPT~\cite{Gil-Dominguez:2023eld}, updated to include the recent data of~\cite{Yeo:2024chk}. The relation between $h_0,h_1$ and the LECs in~\cite{Gil-Dominguez:2023eld} is given in the Supplemental Material~\cite{SM}. We follow the procedure to implement the scattering amplitude between the heavy and light mesons in the finite volume done in~\cite{Gil-Dominguez:2023puj}. See also~\cite{Gil-Dominguez:2024zmr,Zhuang:2024udv}.

\begin{figure*}[!ht]
	\includegraphics[width=0.9\textwidth]{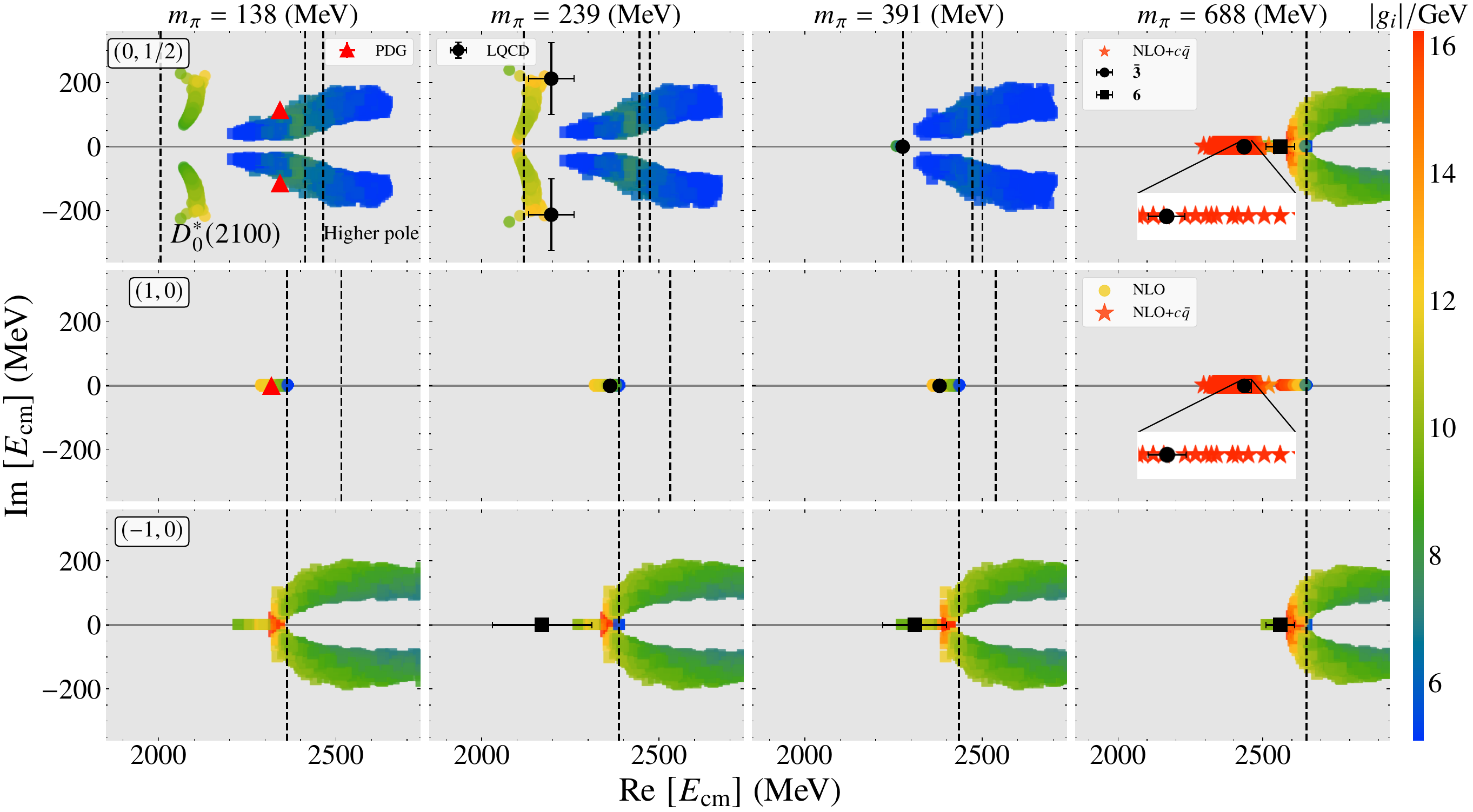}
	\caption{\label{fig:pole-sample-nlo}Pole distributions. Each point corresponds to a pole extracted from one parameter sample. The parameter samples are generated within the $1\sigma$ uncertainty band. The coupling $\abs{g_{i}}$ (in GeV) is encoded in color.}
\end{figure*}

\noindent{\itshape\bfseries Three-pion mass fit.---}%
We perform fits with UChPT at NLO to the energy levels for the open-charm sectors: $(S,I)=(0,\frac{1}{2})$, $(1,0)$, including $D\pi, D\eta, D_s\bar{K}$ and $ DK, D_s\eta$ couple-channel scattering, for the three ensembles with $m_\pi=239$~MeV, $391$~MeV, and $688$~MeV from the LQCD simulations~\cite{Moir:2016srx,Cheung:2020mql,Gayer:2021xzv,Yeo:2024chk} (including the $\mathbf{6}$ and $\overline{\mathbf{15}}$ data from~\cite{Yeo:2024chk}. See the next section for the $\mathbf{\bar{3}}$ analysis.)\footnote{In this fit, we also included the data for the $\mathbf{6}$ and $\overline{\mathbf{15}}$ representations in~\cite{Yeo:2024chk} at $m_\pi=688$ MeV. However, we note that these sets of data do not modify the result obtained with the analysis of data done here from the simulations for $m_\pi=239$~MeV and $391$~MeV~\cite{Moir:2016srx,Cheung:2020mql,Gayer:2021xzv}.}. This corresponds to around $\sim 100$ energy level data points. 
\begin{table}[!htb]
	\renewcommand{\arraystretch}{1.2}
	\setlength{\tabcolsep}{0.25cm}
	\centering
	\caption{\label{tab:fit-LECs}The values of the parameters constrained by the energy levels. The $\chi_\mathrm{d.o.f.}=1.5$. To have dimensionless parameters, we define, $h_{45}=\frac{h_4}{h_5}$, and $h_5' = m_{H,\mathrm{avg}}^2h_5$, where the $m_{H,\mathrm{avg}}$ is the spin average mass of the heavy mesons $D^{(*)}_{(s)}$ in the chiral limit. Here we fix it to $m_{H,\mathrm{avg}}=1929$ MeV.}
	\begin{tabular}{cccc}
		\hline\hline
		$h_2$  & $h_3$ & $h_{45}$ & $h_5'$ \\\hline
		$-0.56(26)(22)$ & $-5.39(5)(35)$ & $0.17(1)(12)$ & $2.6(2)(1)$
		\\
		\hline\hline
	\end{tabular}
\end{table}
We find two $D_0^*$ poles for pion masses below $m_\pi=391$ MeV, one over the $D\pi$ threshold and other one around the $D_s\bar{K}$ threshold. Besides that, we find a bound state related to the $D_{s0}(2317)$. The results for the pole positions and couplings of the bound state related to the $D_{s0}(2317)$ that appears in the $DK$ system are in agreement with the previous analysis of~\cite{Gil-Dominguez:2023puj}, with the result from LQCD~\cite{Cheung:2020mql} and with the experiment~\cite{ParticleDataGroup:2024cfk}. The results for the pole position distribution in the $68\%$ confidence level in comparison with the LQCD and experimental data are displayed in Fig.~\ref{fig:pole-sample-nlo}, where our extrapolation to the physical pion mass is also provided. The central values with their errors are given in Table~\ref{tab:pole-D2300-NLO}. We also show our prediction for the $(S,I)=(-1,0)$ channel ($D\bar{K}$), where the mean value corresponds to a resonance at NLO\footnote{Since here we want to focus in the predictions for the trajectories of the two $D_0^*$ poles, these data are not included in the fit. In addition, we find some tension between the different data sets. In spite of that, our results for the energy levels are in agreement with the analysis of $D\bar{K}$ scattering in~\cite{Cheung:2020mql} as it will be shown latter in this section. }. For the results at LO and in other sectors, see the Supplemental Material~\cite{SM}.  At the physical pion mass, in the sector we are interested, $(S,I)=(0,1/2)$, the pole parameters of the lower $D_0^*$ pole, corresponding to a resonance, are, $\qty[m,\Gamma/2]=\qty[2094(7)(1),111(7)(13)]$~MeV. We call to this pole $D_0^*(2100)$ from now on. The higher pole is located at $\qty[2463(58)(27),108(14)(12)]$ MeV, see Table~\ref{tab:pole-D2300-NLO}. In the $1\sigma$ region this pole can correspond to a virtual state or a resonance in our calculation. The lower resonance becomes a bound state for $m_\pi\simeq 391$~MeV. Note that, here we find the higher energy pole in a different Riemann sheet than in~\cite{Albaladejo:2016lbb}, but the pole parameters are compatible. This is possible since in ~\cite{Albaladejo:2016lbb} a prediction based on the work of~\cite{Liu:2012zya} is made, while here we analyze the data~\cite{Moir:2016srx,Cheung:2020mql,Gayer:2021xzv}. The values of the LECs obtained are given in Table~\ref{tab:fit-LECs}. For the cutoff in the two-meson loop function we obtain $\Lambda=677(45)(50)$~MeV. We find a better agreement with the data when $\Lambda$ is left as free parameter, but we restrict its range around the LO result, which is $\Lambda=795(150)$ MeV. We obtain a non-negligible correlation between the LECs. The correlation matrix is provided in the Supplemental Material~\cite{SM}. In Fig.~\ref{fig:pole-comparsion}, we show our extrapolation for the two $D_0^*$ poles to the physical point in comparison with the experiment (only one resonance) and with other works. Our results for the pole positions are in line with the works of~\cite{Gamermann:2006nm,Guo:2015dha,Albaladejo:2016lbb,Guo:2018tjx, Altenbuchinger:2013vwa}, both poles appearing in the same Riemann sheet as in the work of~\cite{Gamermann:2006nm}, and in disagreement with the PDG average~\cite{ParticleDataGroup:2024cfk} and the result from the lattice simulation close to physical point, i.e., $m_\pi=133$ MeV~\cite{Yan:2024yuq}, that incorporates only $D\pi$ interpolators.\footnote{The higher pole appears as a bump close to the $D\eta$ and $D_s\bar{K}$ thresholds in the $D\pi$ invariant mass distribution, see the Supplemental Material~\cite{SM}, where also the LO results are shown.} The data of~\cite{Yan:2024yuq} are analyzed in~\cite{Luo:2026kui}. 
\begin{figure}[!ht]
    \centering
	\includegraphics[width=0.8\columnwidth]{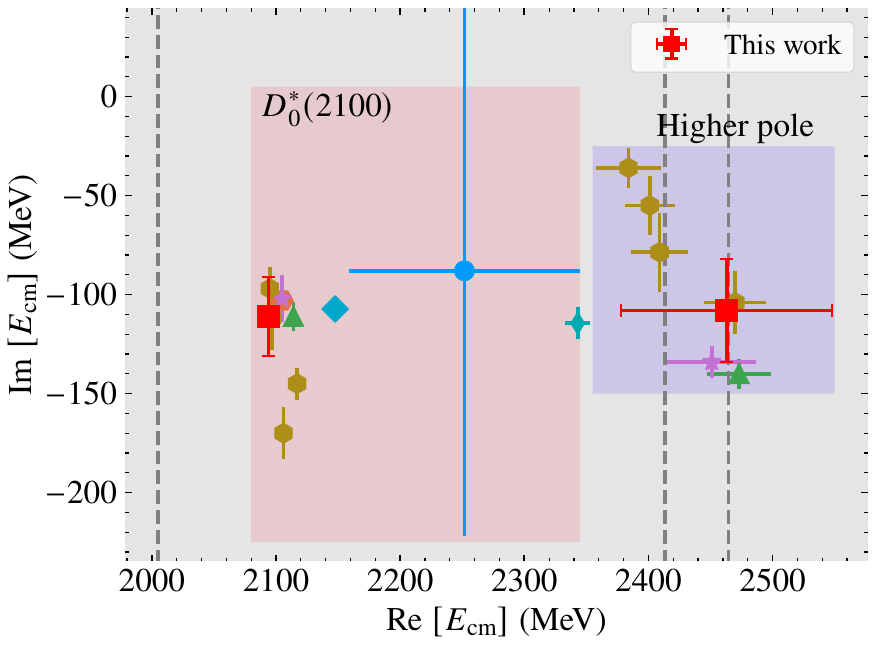}
    \resizebox{0.9\columnwidth}{!}{\input{figure/pole_comparsion_legend.pgf}}
    \caption{\label{fig:pole-comparsion}The positions of the two $D_0^*$ resonances (color in red) up to NLO obtained in this work in comparison with previous works by ChPT~\cite{Altenbuchinger:2013vwa, Guo:2015dha,Albaladejo:2016lbb,Guo:2018tjx, Gamermann:2006nm}, the PDG average~\cite{ParticleDataGroup:2024cfk}, and an analysis of the lattice simulation at one of the pion masses, i.e., $m_\pi=133$ MeV~\cite{Yan:2024yuq} that is very close to physical pion mass. The dashed lines are the thresholds. 
    }
\end{figure}

\begin{figure}
	\centering
	\includegraphics[width=0.9\columnwidth]{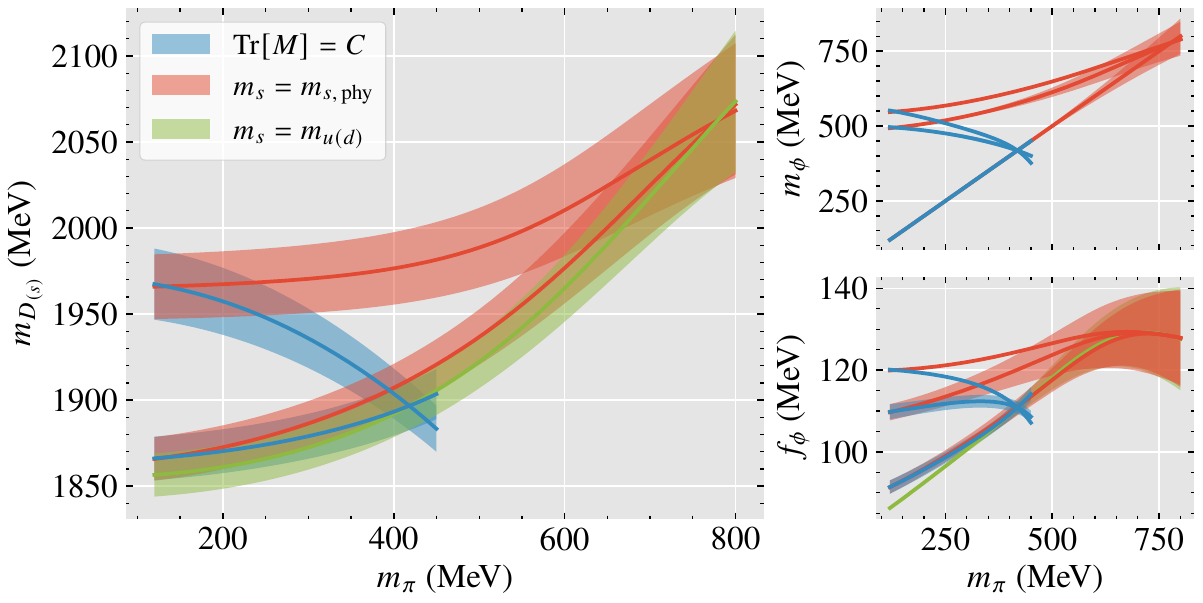}
	\caption{\label{fig:traj-mass-main}The pion mass dependence of the pseudo-scalar meson octet $m_\phi$, the decay constants $f_\phi$, $\phi=\pi,K,\eta$~\cite{deconsnnlo}, and the heavy charm mesons $m_{D_{(s)}}$ under the trajectories $\mathrm{Tr}[M]=C$, $m_s=m_{s,\,\mathrm{phy}}$, and $m_s=m_{u(d)}$, respectively. The charm quark mass is fixed to the physical value. The SU(3) flavor symmetry point is $m_\pi=420$~MeV and $m_\pi=778$ MeV for trajectory $\mathrm{Tr}[M]=C$ and $m_s=m_{s,\,\mathrm{phy}}$, respectively. The mass of the charmed meson with $m_s=m_{s,\mathrm{phy}}$ in the SU(3) flavor symmetry point is $m_H=2062(38)$~MeV.}
\end{figure}

For the first time, the pion mass dependence of the two $D_0^*$ poles in different chiral trajectories is determined. This includes $\mathrm{Tr}[M]=C$\footnote{This trajectory stands for $2m_{ud}+m_s=\mathrm{constant}$, and $m_{ud}$ being the average $m_u, m_d$ mass. }, which encounters the SU(3) symmetric line, $m_{ud}=m_s$ at $m_\pi=420$ MeV, also studied, and the $m_s=m_{s,\mathrm{phy}}$ trajectory, at $m_\pi\simeq 780$ MeV. In Fig.~\ref{fig:traj-mass-main} these trajectories are depicted for the $D_{(s)}$, $\phi=\pi,K,\eta$ mesons in blue, green and red colors, respectively. The masses of the heavy and light mesons as well as the light pseudoscalar decay constants as a function of the pion mass are shown. For the heavy mesons, the analysis of the masses at one-loop NLO of~\cite{Gil-Dominguez:2023eld} is employed, while for the decay constants, we take as input the result from the three-flavor NNLO analysis done recently in~\cite{deconsnnlo}. Note that the latter is necessary since NLO ChPT fails around $m_\pi\simeq 450$~MeV~\cite{Molina:2020qpw,Zhuang:2024udv}.

\begin{figure}[!ht]
	\centering
	\includegraphics[width=0.9\columnwidth]{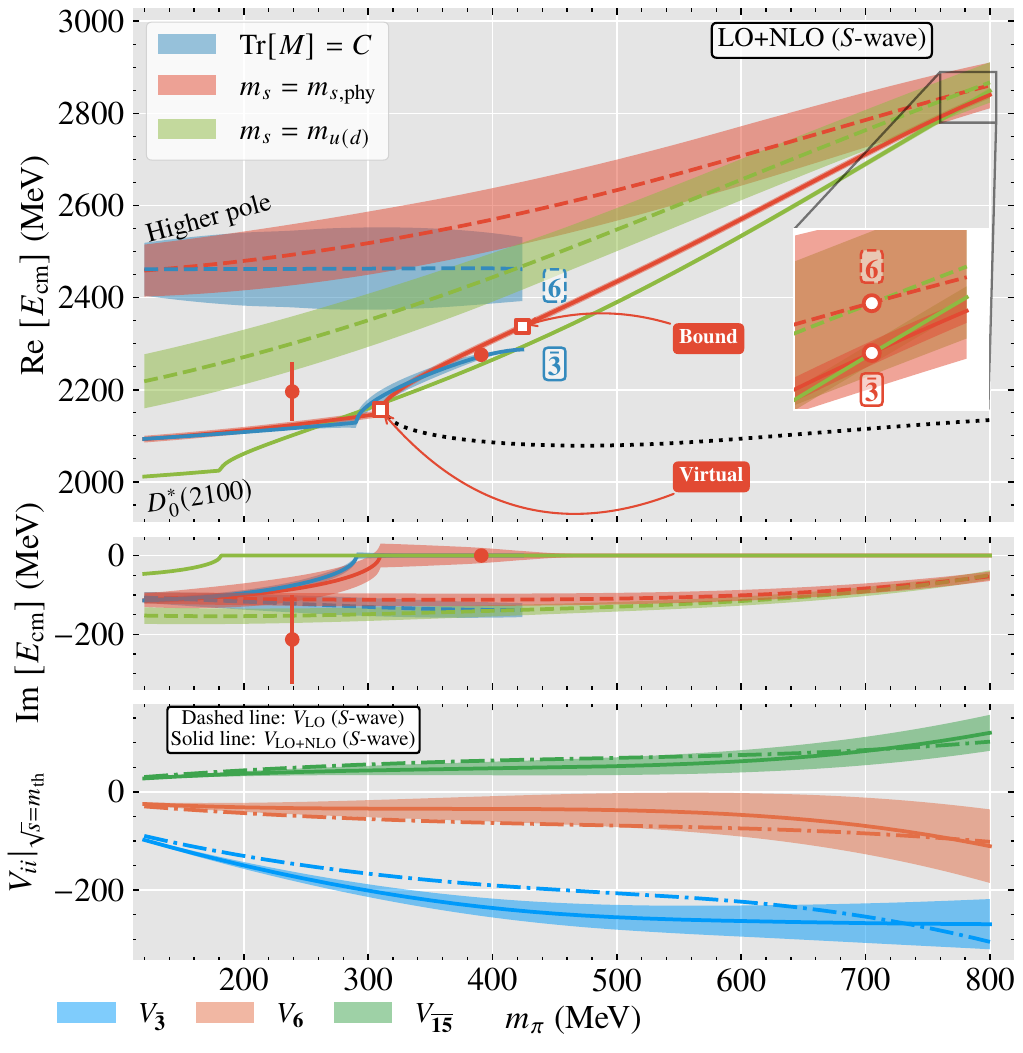}
	\caption{\label{fig:traj-D0-lo-nlo}The trajectories of the two $D_0^*$ poles up to NLO over the chiral trajectories studied, $\mathrm{Tr}[M]=C$, $m_s=m_{s\,\mathrm{phy}}$, and $m_s=m_{u(d)}$, respectively. The error bands represent the $1\text{-}\sigma$ statistical uncertainty. The charm quark mass is fixed to the physical value. The bottom figure shows the diagonal element interactions at the threshold energies as a function of $m_\pi$. LQCD data with error bars for the poles are taken from Refs.~\cite{Moir:2016srx,Gayer:2021xzv}.}
\end{figure}

\begin{figure}[!hbtp]
	\centering
	\includegraphics[width=0.8\columnwidth]{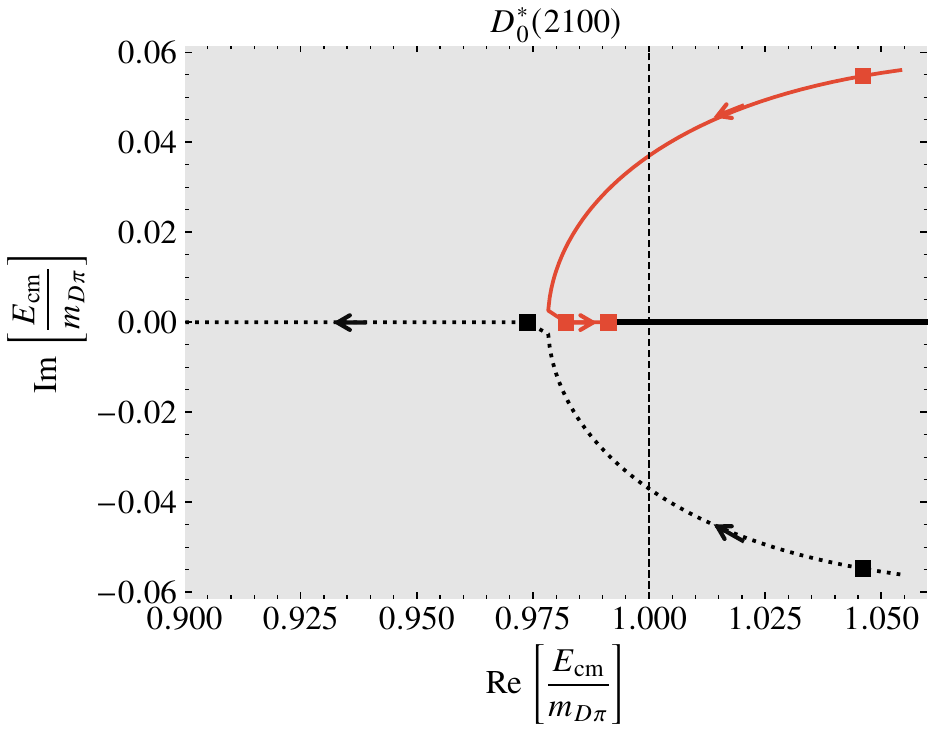}
	\caption{\label{fig:movement-Dpi}Quark mass dependence of the $D_0^*(2100)$ pole relative to the $D\pi$ threshold as a function of $m_\pi$ in the $m_s=m_{s,\mathrm{phy}}$ trajectory. The boxes denote the pole position at $m_\pi=138$, $310$, and $780$~MeV, the physical point, where the pole splits into two and the SU(3) limit, respectively. The black-dot line is the $D\pi$ threshold.}
\end{figure}

The results for the trajectories of the two $D_0^*$ poles at NLO for the physical charm quark mass are depicted in Fig.~\ref{fig:traj-D0-lo-nlo}
The strengths of the interactions as a function of the pion mass for every representation are also displayed. While the interaction for the $\mathbf{\bar{3}}$ is strongly attractive and weekly attractive in the $\mathbf{6}$ representation, it is repulsive for the ${\overline{\mathbf{15}}}$. Hence, no poles can be generated in the ${\overline{\mathbf{15}}}$. On one hand, the lower-energy $D_0^*(2100)$ resonance becomes a virtual state at $m_\pi\simeq 310$~MeV, where it separates into two poles, one that becomes a bound state around $m\simeq 420$ MeV when moving in the $m_s=m_{s,\mathrm{phy}}$ trajectory,\footnote{This small difference with the previous result where it becomes bound at $m_\pi\simeq 391$ MeV is due to the fact that in~\cite{Moir:2016srx} the charm quark mass is slightly lower than the physical one~\cite{Gayer:2021xzv}.} and another virtual bound pole that moves away from the threshold as it is depicted in Fig.~\ref{fig:movement-Dpi}. This behavior is very similar to the one of the $\sigma$ resonance in $\pi\pi$ scattering~\cite{Hanhart:2008mx}. The pion mass dependence of the $D^*_0(2100)$ coupling to $D\pi$ is reported in~\cite{SM} and fully resembles the one for the $\sigma$ previously studied~\cite{Guo:2018zss}. At the SU(3) limit, this pole is bound, and the value of the \textit{compositeness} is meaningful. We obtain that in the UChPT description of $D\pi$ scattering, $X=0.99_{(3)(2)}^{(1)(1)}$, meaning that, in our prediction from the analysis done in this section, the $D_0^*(2100)$ becomes a molecular state in the SU(3) limit. We also obtain that the $D_0^*(2100)$ pole is connected to the $\mathbf{\bar{3}}$ representation. The pole position for the lower energy $D\pi$ resonance is consistent with the LQCD data~\cite{Moir:2016srx,Gayer:2021xzv}, but not with the $\mathbf{\bar{3}}$ data of ~\cite{Yeo:2024chk}, as shown in the picture. The latter will be analyzed in the next section. On the other hand, the $D_0^*(2300)$ high-energy pole remains being a resonance but very close to the threshold till reaching the SU(3) limit. This pole is related to the representation $\mathbf{6}$, in agreement with previous works~\cite{Gamermann:2006nm,Albaladejo:2016lbb}. As a consequence, this pole cannot be affected by a possible $Q\bar{q}$ state in this limit. Note that in~\cite{Godfrey:1985xj} the $Q\bar{q}$ state is predicted to have a mass around $2.4$~GeV at the physical pion mass. While in~\cite{Albaladejo:2016lbb} the high-energy resonance evolves from the $[--+]$ sheet at the physical point to the $[---]$ sheet near the symmetric point, here it is found for all pion masses studied at the $[---]$ sheet, in agreement with~\cite{Gamermann:2006nm}\footnote{In the SU(3) limit, all the diagonal elements of the $G$ function loop must be identical according to the SU(3) symmetry, and therefore, there are only two Riemann sheets in this limit, that we simply denote as $[+]$ and $[-]$. See Appendix~A in~\cite{Jido:2003cb}.}. In this work, we have accounted for the pion mass dependence from different sources, hence, providing more accurate predictions than in previous analyses. 

The pole positions at the SU(3) limit are given in Table~\ref{tab:pole-su3-nlo} for the NLO. Different chiral trajectories are also investigated, and, as a result, we observe that the pion mass dependence of the higher pole in the $\mathrm{Tr}[M]=C$ trajectory is approximately constant due to its hidden strange content, coming from the coupling to the $D\eta$ and $D_s\bar{K}$ channels. In contrast, the trajectories of the \(D_0^*(2100)\) over the \(m_s=m_{s,\mathrm{phy}}\) and \(\mathrm{Tr}[M]=C\) lines are compatible up to $m_\pi\simeq 420$~MeV, indicating that the \(s\bar{s}\) component in \(D_0^*(2100)\) is small. This observation, consequence of the predictions made in this work, can be tested in LQCD simulations, and it could be a strong indication of the two $D_0^*$-pole hypothesis. 
\begin{table*}[!hbtp]
	\renewcommand{\arraystretch}{1.8}
	\setlength{\tabcolsep}{0.56cm}
	\centering
	\caption{\label{tab:pole-D2300-NLO}The pole positions of the two $D_0^*$ resonances and their couplings to the channels.}
	\begin{tabular}{cccccc}
		\hline\hline
		$m_\pi$~(MeV) & R.S. & Pole~(MeV) & $\abs{g_{D\pi}}$~(GeV) & $\abs{g_{D\eta}}$~(GeV) & $\abs{g_{D_s\bar{K}}}$~(GeV) \\\hline
		\multirow{2}{*}{$138$} & $[-++]$ & $2094(7)(1)-i111(7)(13)$ & $9.6(2)(2)$ & $2.0(2)(3)$ & $6.8(5)(3)$ \\
		~ 					   & $[---]$ & $2463(58)(27)-i108(14)(12)$ & $4.4(1)(6)$ & $6.3(1)(5)$ & $6.1(4)(3)$ \\\hline
		\multirow{2}{*}{$239$} & $[-++]$ & $2125(7)(1)-i84(9)(17)$ & $10.6(1)(1)$ & $2.5(2)(4)$ & $7.8(5)(3)$ \\
		~ 					   & $[---]$ & $2492(66)(19)-i113(16)(9)$ & $4.7(2)(5)$ & $6.1(2)(3)$ & $5.9(2)(4)$ \\\hline
		\multirow{2}{*}{$391$} & $[+++]$ & $2277(7)(4)$ & $2.3(1)(16)$ & $0.6(4)(1)$ & $1.8(8)(5)$ \\
		~                      & $[---]$ & $2549(68)(15)-i121(16)(7)$ & $4.9(1)(4)$ & $5.8(1)(3)$ & $5.3(2)(2)$ \\\hline\hline
	\end{tabular}
\end{table*}
\begin{table*}[!hbt]
	\centering
	\caption{\label{tab:pole-su3-nlo}The poles~(MeV) and the couplings~(GeV) in the SU(3).}
	\renewcommand{\arraystretch}{2.}
	\setlength{\tabcolsep}{0.5cm}
	\begin{tabular}{cccccc}\hline\hline
		$m_\pi$~(MeV) & R.S. & Threshold~(MeV) & Pole~(MeV) & $\abs{g_\mathbf{\bar{3}}}$~(GeV) & $\abs{g_\mathbf{6}}$~(GeV) \\\hline
		\multirow{2}{*}{$420$~($\mathrm{Tr}[M]=C$)} & $[+]$ & \multirow{2}{*}{\(2317\)} & $2286(2)(13)$  & $10.7(3)(10)$ & $0$ \\
		~ 											& $[-]$ & ~ & $2462(70)(10)-i137(18)(3)$ & $0$ & $9.1(2)(2)$ \\\hline
		\multirow{3}{*}{$688$} & \multirow{2}{*}{$[+]$} & \multirow{3}{*}{\(2650\)} & $2612(6)(14)$~(NLO) & $13.2(5)(20)$ & $0$ \\
		~                      & ~ & ~ & $2443(25)(1)$~$(\mathrm{NLO}+c\bar{q})$ & $16.8(4)(5)$ & $0$ \\
		~                      & $[-]$ & ~ & $2735_{(40)(180)}^{(60)(120)}-i107_{(50)(60)}^{(60)(10)}$ & $0$ & $9.2(3)(4)$ \\\hline
		\multirow{2}{*}{$780$} & $[+]$ & \multirow{2}{*}{\(2841\)} & $2816(7)(11)$ & $13.0(5)(27)$ & $0$\\
		~                      & $[-]$ & ~ & $2846(50)(5)-i61_{(30)(30)}^{(10)(30)}$ & $0$ & $10.5(3)(10)$\\
		\hline\hline
	\end{tabular}
\end{table*}

\begin{table}
	\renewcommand{\arraystretch}{1.3}
	\setlength{\tabcolsep}{0.35cm}
	\centering
	\caption{\label{tab:qqbar-parameter}The couplings and $m_{c\bar{s}}$ in Eq~\eqref{eq:Vcsbar} constrained by energy levels~\cite{Moir:2016srx,Yeo:2024chk,Gayer:2021xzv,Cheung:2020mql}.}
	\begin{tabular}{cccc}
		\hline\hline
		~ & $g_0$~(GeV) & $g_1$ & $m_{c\bar{s}}$~(MeV)\\\cline{2-4}
		LO + $c\bar{q}$ & $4.9(10)$ & $40(13)$ & $2550(96)$ \\
		NLO + $c\bar{q}$ & $2.4(5)(5)$ & $40(6)(7)$ & $2550$~[Fixed]\\\hline\hline
	\end{tabular}
\end{table}
\noindent{\itshape\bfseries Fit to the SU(3) data.---}%
As mentioned, we did not include the $\mathbf{\bar{3}}$ data for $m_\pi\simeq 688$~MeV~\cite{Yeo:2024chk} in the previous section. The reason is twofold. On one hand, the SU(3) simulation of~\cite{Yeo:2024chk} for $m_\pi\simeq 688$ MeV entails larger systematic errors than the lower pion mass ones~\cite{Moir:2016srx,Cheung:2020mql,Gayer:2021xzv}, that can be due mostly to a larger temporal and spacial lattice spacing, see~\cite{SM} and the discussion in~\cite{Yeo:2024chk}, leading to some tension in combined fits with data from the other two LQCD simulations. Note also that, as shown in Fig.~\ref{fig:traj-mass-main}, in our fit to the charmed meson masses available~\cite{Gil-Dominguez:2023eld}, the symmetric point in the $m_s=m_{s,\mathrm{phy}}$ trajectory is reached at a pion mass of $780$ MeV, differing from the one of~\cite{Yeo:2024chk} ($688$ MeV). Still, fixing the charm quark mass to the physical value in LQCD simulations entails larger systematic errors as compared to the binding energies of some of the two-meson molecules close to threshold as the $D_{s0}(2317)$~\cite{Gil-Dominguez:2023puj}. On the other hand, $Q\bar{q}$ is in the $\mathbf{\bar{3}}$ representation. Indeed, when inspecting carefully the results from~\cite{Yeo:2024chk}, we observe some essential features. First, in the SU(3) simulation of~\cite{Yeo:2024chk}, two poles corresponding to bound states are extracted from the analysis, one that connects to the $\mathbf{\bar{3}}$ representation, and the other one to the $\mathbf{6}$, while only one pole was found in~\cite{Moir:2016srx,Cheung:2020mql,Gayer:2021xzv}. The former being consistent with the two-$D_0^*$ pole hypothesis from UChPT. Second, the fact that the interaction is strongly attractive for the $\mathbf{\bar{3}}$, weakly attractive for the $\mathbf{6}$, and repulsive in the $\mathbf{\overline{\mathbf{15}}}$ is also consistent qualitatively with the UChPT findings. Third, while in~\cite{Moir:2016srx,Cheung:2020mql,Gayer:2021xzv} the pole extracted related to the $D_0^*(2100)$ is close to the free energy level for $D\pi$ scattering, and far from the first energy level related to $Q\bar{q}$, the lowest pole extracted in~\cite{Yeo:2024chk} is close to this $Q\bar{q}$ energy level. Indeed, we note that higher quality fits are obtained when a bare $Q\bar{q}$ pole (CQM) is included in the interaction, coupling to the heavy-light two-meson channel.  For these reasons, here we investigate separately the inclusion of the $\bar{3}$ data of~\cite{Yeo:2024chk} by conducting combined fits of the \cite{Moir:2016srx,Cheung:2020mql,Gayer:2021xzv,Yeo:2024chk} data with, (i) only NLO, (ii) LO+$c\bar{q}$, and (iii) NLO+$c\bar{q}$. To take into account the bare component, we consider a CDD pole term, as follows, 
\begin{equation}\label{eq:vt}
V_\mathrm{tot}=V+V_{c\bar{q}}\ ,
\end{equation}
where $V$ stands for the LO or NLO heavy-light two-meson interaction in UChPT, and the bare term is given by,
\begin{equation}\label{eq:Vcsbar}
	V_{c\bar{q}} = \frac{g^2}{s - m_{c\bar{q}}^2}. 
\end{equation}
In the $m_s=m_{s,\mathrm{phy}}$ trajectory and in the SU(3) limit, we denote to the bare mass, $m_{c\bar{q}}^{0}$, that indeed corresponds to $m_{c\bar{s}}$. To incorporate a possible $g$-coupling light-quark mass dependence, we note that a simple first-order Taylor expansion in $m_\pi$, such as,
\begin{equation}\label{eq:qqbar-coupling}
	g = g_0 + (m_\pi - m_{\pi,\mathrm{phy}}) g_1,
\end{equation}
 works better compared to an expansion in $m_\pi^2$. While for the $m_{c\bar{q}}$ we take, 
\begin{equation}\label{eq:cqbar}
	m_{c\bar{q}}=m_D+m_{c\bar{q}}^{0}-m_{D}^{0}\ ,
\end{equation}
where, $m_{c\bar{q}}^{0}$, $m_{D}^{0}$, correspond to the values of the bare mass and $D$ meson mass in the SU(3) limit for the trajectory  $m_s=m_{s,\mathrm{phy}}$. Hence, $m_{c\bar{q}}^{0}=m_{c\bar{s}}$, and, for the $m_{D}^{0}$, we take the average value between the result of our fit of the heavy meson masses~\cite{Gil-Dominguez:2023eld} and the one of the LQCD simulation~\cite{Yeo:2024chk}, i. e., $m_{D}^{0}=(1962+2062)/2$, taking into account the systematic uncertainty as the difference between the two values in the error bands plotted. We also notice that a small variation of this number does not affect the $\chi^2$ of the fit.

The result for the CDD pole parameters are yielded in Table~\ref{tab:qqbar-parameter}. In the NLO fit, the value of $m_{c\bar{s}}$ is fixed to the LO result. The result for the bare mass is in agreement with the quark model prediction~\cite{Godfrey:1985xj}. While other sectors are not affected compared with the results obtained in the previous section, or in the physical region for the $D_0^*(2100)$, overall, the results obtained related to the $\mathbf{\bar{3}}$ around the SU(3) limit now describe well the findings in~\cite{Yeo:2024chk}.
This can be seen in Fig.~\ref{fig:pole-sample-nlo}, where the pole position of the $D_0^*(2100)$ is reduced by around $170$ MeV compared to the NLO result without the $c\bar{q}$, being consistent with the value reported in~\cite{Yeo:2024chk} within the $1\sigma$ uncertainty. The description of the data with NLO$+c\bar{q}$ significantly improves the quality of the fit compared to using the LO$+c\bar{q}$ parameterization. The results for the energy levels and phase shifts are provided in the Supplemental material~\cite{SM}. With the NLO$+c\bar{q}$ result, we evaluate the \textit{compositeness} of the $D_0^*(2100)$ in the SU(3) limit. We obtain now $X=0.63(11)(26)$, which is considerably reduced compared with the result in the previous section. This value is now compatible with the estimation from ~\cite{Yeo:2024chk}, $0.56(2)(10)$.~\footnote{We estimate the \textit{compositeness} of the pole in $\mathbf{\bar{3}}$ sector~\cite{Yeo:2024chk} by employing the relation, $X=\qty(1+2\abs{r_0/a_0})^{-1/2}$~\cite{Matuschek:2020gqe}, where the scattering length $a$ and the effective range $r$ values are $a_0=-a_t 8.54(31)(64)$, and $r_0=-a_t 9.46(86)(40.7)$. }
This result, which should be tested in other LQCD simulations, indicates that the coupling of the $D_0^*(2100)$ to the $c\bar{q}$ increases significantly near the SU(3) limit, while for pion masses near the physical value the molecular description provided by the UChPT two-meson scattering analysis can describe well the experimental and LQCD data.~\footnote{When including the bare component, its value around the physical pion mass is $m_{c\bar{q}}=2400(32)(60)$~MeV, consistently with~\cite{Godfrey:1985xj}. However, the fit result for the $D_0^*(2100)$ pole position barely changes in the physical point compared to the NLO UChPT description.}

\noindent{\itshape\bfseries Conclusion.---}%
We have preformed an exhaustive analysis of the scattering LQCD data of a light-pseudoscalar meson with a charmed one within NLO UChPT. For the first time, the pion mass dependence for various chiral trajectories of the two poles related to the experimental $D_0^*(2300)$ are investigated, $m_s=m_{s,\mathrm{phy}}$, $\mathrm{Tr}[M]=C$, and $m_s=m_{u(d)}$. We observe that the lower pole separates into two poles as the pion mass increases, a virtual bound and a bound state, coupling to the $\bar{\mathbf{3}}$ representation at the SU(3) limit. While the UChPT analysis provides a compositeness $X\simeq 1$ at this limit, the SU(3) LQCD data clearly demands for the inclusion of a CDD pole or genuine $Q\bar{q}$ component at these high pion masses, $m_\pi\simeq 700$~MeV, being it reduced to $X=0.63(11)(26)$. This phenomena should be investigated in future LQCD simulations. On the other hand, the higher pole connected to the $D_0(2300)$ can be a resonance or virtual state in the $1\sigma$ region, and it is connected to the representation $\mathbf{6}$ in the SU(3) limit. We find that the real part of this pole does not increase with the pion mass in the $\mathrm{Tr}[M]=C$ trajectory due to its hidden strange content, i.e. coupling to the $D\eta, D_{s}\bar{K}$ channels. If tested by future LQCD simulations, this finding can be a relevant evidence for the existence and nature of this pole. Finally, a bound state in the strange sector $S=1$ coupling to the $\bar{\mathbf{3}}$ representation and a resonance that can become virtual in the $S=-1$ sector and in the $\mathbf{6}$ representation at the SU(3) limit, are found. As we have shown, our results are compatible with the LQCD findings and provide support for the higher pole of the $D_0^*(2300)$ state. 

\noindent{\itshape\bfseries Acknowledgments.---}%
Z. Zhuang and R. Molina are grateful to N. Lang, Julian~A.~S., E. Oset, Fang-Zheng~Peng, Haobo~Yan, and Yao~Ma for useful discussions. We acknowledge to the Hadron-Spectrum Collaboration for providing the data. The plots were made with \texttt{Matplotlib}~\cite{Hunter:2007} and the calculations were performed by \texttt{Julia}~\cite{Julia-2017}. 
The \texttt{Optimizer.jl} package~\cite{Optimize} is used to find the finite-volume spectra of the scattering amplitude. This work is supported by the Ministerio de Ciencia e Innovación, research contract PID2023-147458NB-C21,
funded by MICIU/AEI/10.13039/501100011033. R. M. also acknowledges support from the ESGENT program with Ref. ESGENT/018/2024 and the PROMETEU program with Ref. CIPROM/2023/59, of the Generalidad Valenciana (GVA). This project has received funding from the European Union Horizon 2020 research
and innovation program under the program H2020-INFRAIA-2018-1, grant agreement No. 824093 of the STRONG-2020.
%
%
%

\clearpage
\newpage
\setcounter{page}{1}
\setcounter{figure}{0}  
\setcounter{table}{0}  
\setcounter{equation}{0}
\begin{widetext}

\section{Supplemental Material}
In this supplemental material, we present details regarding the Formalism, the pion mass dependence of the heavy \(D\) mesons, the summary of lattice setups, the fitted energy levels, the correlations between the fitted parameters, the pion mass dependence of the poles in some sectors, as well as the predicted scattering lengths, and phase shifts.
\subsection{Formalism}
The leading order Lagrangian describing the interactions between the charmed-meson triplet $\mathcal{P}=\qty(D^0, D^+, D_s^+)$ and pseudoscalar meson octet reads
	\begin{equation}\label{eq:lag_lo}
		\mathcal{L}_{\mathcal{P}\phi}^{(1)} =
		\covpd{\mathcal{P}}{\mu}\covpd{\mathcal{P}^\dagger}{\mu} - m_{0H}^2\mathcal{P}\mathcal{P}^\dagger 
	\end{equation}
where the covariant derivative $\covpd{\mathcal{P}}{\mu}:=\mathcal{P}\qty(\overleftarrow{\partial}_\mu+\Gamma_\mu^\dagger)$. $m_{0H}$ is a charmed-meson mass in the chiral limit. The Lagrangian at next to leading order reads~\cite{Guo:2008gp,Meng:2022ozq,Guo:2009ct,Yao:2015qia}
	\begin{equation}\label{eq:lag_nlo}
		\mathcal{L}^{(2)}_{\mathcal{P}\phi} = \mathcal{P}\qty(-h_0\Tr[\chi_+]-h_1\chi_+ +h_2\Tr[u_\mu u^\mu]-h_3u_\mu u^\mu)\mathcal{P}^\dagger +\covpd{\mathcal{P}}{\mu}\qty(h_4\Tr[u_\mu u^\nu]-h_5\anticom{u^\mu}{u^\nu})\covpd{\mathcal{P}^\dagger}{\nu}.
	\end{equation}
The vector current $\Gamma_\mu$ and axial current $u_\mu$ are given by
\begin{equation}\label{eq:vec_current}
	\Gamma_\mu = \frac{1}{2}\qty(u^\dagger\pd{u}{\mu}+u\pd{u^\dagger}{\mu}) = \frac{1}{2}\com{u^\dagger}{\pd{u}{\mu}},\footnote{Note that \(u^\dagger\pd{u}{\mu}+u\pd{u^\dagger}{\mu}=u^\dagger\pd{u}{\mu}-\pd{u}{\mu}u^\dagger=\qty[u^\dagger,\pd{u}{\mu}]\).}
\end{equation}
and 
\begin{equation}\label{eq:axial_current}
	u_\mu = i\qty(u^\dagger\pd{u}{\mu}-u\pd{u^\dagger}{\mu}),
\end{equation}
with $u^2 = U=\exp\qty(i\frac{\sqrt{2}\phi}{f})$. The field $\chi_+$ reads
\begin{equation}
	\chi_+ = u^\dagger\chi u^\dagger + u\chi^\dagger u,
\end{equation}
with $\chi=2B_0s$, where $s$ denotes the scalar external sources. Using the Gell-Mann-Oakes-Renner relations, the $\chi$ is rewritten as $\chi=\mathrm{diag}(m_{K^+}^2-m_{K^0}^2+m_{\pi^0}^2,m_{K^0}^2-m_{K^+}^2+m_{\pi^0}^2,m_{K^+}^2+m_{k^0}^2-m_{\pi^0}^2)$. 
We have set the all external currents to zero in the Eqs.~\eqref{eq:vec_current} and \eqref{eq:axial_current}. The pseudo-scalar meson field $\phi$ is given by
\begin{equation}
	\phi = \begin{pmatrix}
		\frac{\pi^0}{\sqrt{2}}+\frac{\eta}{\sqrt{6}} & \pi^+ & K^+ \\
		\pi^- & -\frac{\pi^0}{\sqrt{2}}+\frac{\eta}{\sqrt{6}} & K^0 \\
		K^- & \bar{K}^0 & -\frac{2}{\sqrt{6}}\eta
	\end{pmatrix}
\end{equation} 
\begin{figure}[!h]
	\centering
	\includegraphics[width=0.4\textwidth]{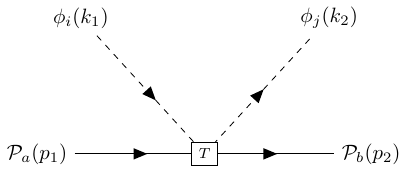}
	\caption{\label{fig:scattering-process}Diagrammatic representation for the $\mathcal{P}_a(p_1)\phi_i(k_1)\to \mathcal{P}_b(p_2)\phi_j(k_2)$ process. }
\end{figure}
For a process $\mathcal{P}_a(p_1)\phi_i(k_1)\to \mathcal{P}_b(p_2)\phi_j(k_2)$, see Fig.~\ref{fig:scattering-process}, Eq.~\eqref{eq:lag_lo} generates the interaction kernel called Weinberg-Tomozawa term, i.e.
\begin{equation}
	V_\mathrm{WT} = -\frac{C_\mathrm{LO}^{(ia,jb)}}{4f_i f_j}(s-u),
\end{equation}
with the Mandelstam variables $s$ and $u$. The coefficients $C_{ia,jb}$ with different strangeness and isospin are list in Table~\ref{tab:coeff_isospin_space}. The NLO interaction kernel extracted from the Eq.~\eqref{eq:lag_nlo} has the form,
\begin{equation}
	V_\mathrm{NLO} = \frac{-1}{f_if_j}\left(C_{h_0}^{(ia,jb)}h_0+C_{h_1}^{(ia,jb)}h_1+H_{24}(s,t,u)+H_{35}(s,t,u)\right),
\end{equation}
where $H_{24}(s,t,u)$ and $H_{35}(s,t,u)$ are given by
\begin{equation}\label{eq:H24}
	H_{24}\qty(s,t,u) = C_{h_2}^{(ia,jb)}h_2\qty(k_1\cdot k_2)+C_{h_4}^{(ia,jb)}h_4\qty[(k_1\cdot p_2)(k_2\cdot p_1)+(k_1\cdot p_1)(k_2\cdot p_2)],
\end{equation}

and
\begin{equation}\label{eq:H35}
	H_{35}\qty(s,t,u) = C_{h_{3(5)}}^{(ia,jb)}\left[h_3(k_1\cdot k_2)+h_5(k_1\cdot p_2)(k_2\cdot p_1)+(k_1\cdot p_1)(k_2\cdot p_2)\right]	,
\end{equation}
respectively. The LECs $h_l$, $l=0,3$ are dimensionless, while $h_{4(5)}$ have dimensions of $\mathrm{Mass}^{-2}$. In Eq.~\eqref{eq:H24}, the coefficients $C_{h_2}^{(ia,jb)}$ and $C_{h_4}^{(ia,jb)}$ satisfy the relation $C_{h_2}^{(ia,jb)}=2C_{h_4}^{(ia,jb)}$. The coefficients $C_{h_l}^{(ia,jb)}$, $l=0,5$ are list in Table~\ref{tab:coeff_isospin_space}.

\begin{table*}[!hbtp]
	\renewcommand{\arraystretch}{2.5}
	\setlength{\tabcolsep}{0.2cm}
	\centering
	\caption{\label{tab:coeff_isospin_space}The coefficients of the process $\mathcal{P}_a\phi_i\to\mathcal{P}_b\phi_j$ in the interaction kernels. The phase conventions used for channel space are taken from Ref.~\cite{Gamermann:2006nm}. The $\mu_1^2$, $\mu_2^2$, $\mu_3^2$ and the $\mu_4^2$ are defined as $\mu_1^2=4m_K^2-m_\pi^2$, $\mu_2^2=mK^2+m_\pi^2$, $\mu_3^2=5m_K^2-3m_\pi^2$, and $\mu_4^2 = 2m_K^2-m_\pi^2$, respectively. Note that $C_{h_2}=2C_{h_4}$.}
	\begin{tabular}{lccccccc}
	\hline\hline
	~ & $S=0, I=\frac{1}{2}$ & $S=1,I=0$ & $S=1,I=1$ & $S=-1,I=0$ & $S=-1,I=1$ & $S=2,I=1/2$ & $S=0,I=3/2$\\\cline{2-8}
	$C_\mathrm{LO}$ & $\begin{pmatrix}
			 2 & 0 & \sqrt{\frac{3}{2}} \\
 			0 & 0 & \sqrt{\frac{3}{2}} \\
 			\sqrt{\frac{3}{2}} & \sqrt{\frac{3}{2}} & 1
			\end{pmatrix}$ & $\begin{pmatrix}
				2 & \sqrt{3} \\
				 \sqrt{3} & 0 \\
			\end{pmatrix}$ & $\begin{pmatrix}
				0 & 1 \\ 1 & 0
			\end{pmatrix}$ & $1$ & $-1$ & $-1$ & $-1$\\
	$C_{h_0}$ & $\begin{pmatrix}
					4m_\pi^2 & 0 & 0 \\
					0 & \frac{4}{3}\mu_1^2 & 0 \\
					0 & 0 & 4m_K^2
				\end{pmatrix}$ & $\begin{pmatrix}
					4m_K^2 & 0 \\
					0 & \frac{4}{3}\mu_1^2
				\end{pmatrix}$ & $\begin{pmatrix}
					4m_\pi^2 & 0 \\ 0 & 4m_K^2
				\end{pmatrix}$ & $4m_K^2$ & $4m_K^2$ & $4m_K^2$ & $4m_\pi^2$\\
	$C_{h_1}$ & $\begin{pmatrix}
		2m_\pi^2 & 2m_\pi^2 & \sqrt{\frac32}\mu_2^2 \\
					2m_\pi^2 & \frac{2m_\pi^2}{3} & -\frac{\mu_3^2}{\sqrt{6}} \\
					\sqrt{\frac{3}{2}}\mu_2^2 & -\frac{\mu_3^2}{\sqrt{6}} & 2m_K^2
	\end{pmatrix}$ & $\begin{pmatrix}
		4m_K^2 & \frac{\mu_3^2}{\sqrt{3}} \\
					\frac{\mu_3^2}{\sqrt{3}} & \frac83\mu_4^2
	\end{pmatrix}$ & $\begin{pmatrix}
		0 & -\mu_2^2 \\ -\mu_2^2 & 0
	\end{pmatrix}$ & $-2m_K^2$ & $2m_K^2$ & $2m_K^2$ & $2m_\pi^2$\\			
	$C_{h_2}$ & $\begin{pmatrix}
					4 & 0 & 0 \\
					0 & 4 & 0 \\
					0 & 0 & 4
				\end{pmatrix}$ & $\begin{pmatrix}
					4 & 0 \\
					0  & 4
				\end{pmatrix}$ & $\begin{pmatrix}
					4 & 0 \\ 0 & 4 
				\end{pmatrix}$ & $4$ & $4$ & $4$ & $4$\\
	$C_{h_{3(5)}}$ & $\begin{pmatrix}
		-2 & -2 & -\sqrt{6} \\
					 -2 & -\frac{2}{3} & \sqrt{\frac{2}{3}} \\
					 -\sqrt{6} & \sqrt{\frac{2}{3}} & -2 \\
				\end{pmatrix}$ & $\begin{pmatrix}
					 					-4 & -\frac{2}{\sqrt{3}} \\
 					-\frac{2}{\sqrt{3}} & -\frac{8}{3}
				\end{pmatrix}$ & $\begin{pmatrix}
					0 & 2 \\ 2 & 0
				\end{pmatrix}$ & $2$ & $-2$ & $-2$ & $-2$\\\hline\hline
	\end{tabular}
\end{table*}
In the limit of exact SU(3) flavor symmetry, the interaction terms can be decomposed into irreducible presentations. For a \(\phi-\mathcal{P}\) system, the corresponding multiplets are \(\bar{\mathbf{3}},\mathbf{6},\overline{\mathbf{15}}\). The projections of the \(\phi\mathcal{P}\) states in coupled-channel space in the sectors \((0,1/2)\), \((1,0)\), and \((1,1)\) to their relevant multiplets are given by
\begin{equation}\label{eq:multi-S0I12}
	\renewcommand{\arraystretch}{2.2}
	\begin{pmatrix}
		\ket{\bar{\mathbf{3}}} \\
		\ket{\mathbf{6}} \\
		\ket{\overline{\mathbf{15}}}
	\end{pmatrix} = 
	\begin{pmatrix}
		-\frac{3}{4} & -\frac14 & -\sqrt{\frac38} \\
		\sqrt{\frac{3}{8}} & -\sqrt{\frac38} & -\frac12 \\
		\frac14 & \frac34 & -\sqrt{\frac38}
	\end{pmatrix}
	\begin{pmatrix}
		\ket{D\pi} \\
		\ket{D\eta} \\
		\ket{D_s\bar{K}}
	\end{pmatrix},
\end{equation}
\begin{equation}
	\renewcommand{\arraystretch}{2.2}
	\begin{pmatrix}
		\ket{\bar{\mathbf{3}}} \\
		\ket{\overline{\mathbf{15}}}
	\end{pmatrix} = 
	\begin{pmatrix}
		 \frac{\sqrt{3}}{2} & \frac{1}{2} \\
 	\frac{1}{2} & -\frac{\sqrt{3}}{2}
	\end{pmatrix}
	\begin{pmatrix}
		\ket{DK} \\
		\ket{D_s\eta}
	\end{pmatrix},
\end{equation}
and
\begin{equation}
	\renewcommand{\arraystretch}{2.2}
	\begin{pmatrix}
		\ket{\mathbf{6}} \\
		\ket{\overline{\mathbf{15}}}
	\end{pmatrix} = 
	\begin{pmatrix}
		 \frac{1}{\sqrt{2}} & \frac{1}{\sqrt{2}} \\
 		-\frac{1}{\sqrt{2}} & \frac{1}{\sqrt{2}}
	\end{pmatrix}
	\begin{pmatrix}
		\ket{D_s\pi} \\
		\ket{DK}
	\end{pmatrix}.
\end{equation}
The sector \((-1,0)\) couples to the \(\mathbf{6}\), and the sectors \((0,3/2)\), \((2,1/2)\), and \((-1,1)\) to the \(\overline{\mathbf{15}}\). The corresponding coefficients are listed in Table~\ref{tab:coeff_su3}.
\begin{table*}[!htpb]
	\renewcommand{\arraystretch}{2.5}
	\setlength{\tabcolsep}{0.2cm} 
	\centering
	\caption{\label{tab:coeff_su3}The coefficients of the interaction kernels in the SU3 symmetry. $m_\phi$ stands for the meson octet mass in the SU3 symmetry.}
	\begin{tabular}{lccccccc}\hline\hline
		~ & $S=0, I=\frac{1}{2}$ & $S=1,I=0$ & $S=1,I=1$ & $S=-1,I=0$ & $S=-1,I=1$ & $S=2,I=1/2$ & $S=0,I=3/2$\\
		~ & $\{\mathbf{\bar{3}},\mathbf{6},\mathbf{\overline{15}}\}$ & $\{\mathbf{\bar{3}},\mathbf{\overline{15}}\}$ & $\{\mathbf{6},\mathbf{\overline{15}}\}$ & $\{\mathbf{6}\}$ & $\{\mathbf{\overline{15}}\}$ & $\{\mathbf{\overline{15}}\}$ & $\{\mathbf{\overline{15}}\}$\\\cline{2-8}
		$C_\mathrm{LO}$ & $\begin{pmatrix}
			3 & 0 & 0\\
			0 & 1 & 0 \\
			0 & 0 & -1
		\end{pmatrix}$ & $\begin{pmatrix}
			3 & 0 \\0 & -1
		\end{pmatrix}$ & $\begin{pmatrix}
			1 & 0\\0&-1
		\end{pmatrix}$ & $1$ & $-1$ & $-1$ & $-1$\\
		$C_{h_0}$ & $\begin{pmatrix}
			4m_\phi^2 & 0 & 0 \\
			0 & 4m_\phi^2 & 0 \\
			0 & 0 & 4m_\phi^2
		\end{pmatrix}$ & $\begin{pmatrix}
			4m_\phi^2 & 0 \\0 & 4m_\phi^2
		\end{pmatrix}$ & $\begin{pmatrix}
			4m_\phi^2 & 0 \\0 & 4m_\phi^2
		\end{pmatrix}$ & $4m_\phi^2$ & $4m_\phi^2$ & $4m_\phi^2$ & $4m_\phi^2$\\
		$C_{h_1}$ & $\begin{pmatrix}
			\frac{14m_\phi^2}{3} & 0 & 0 \\
			0 & -2m_\phi^2 & 0 \\
			0 & 0 & 2m_\phi^2
		\end{pmatrix}$ & $\begin{pmatrix}
			\frac{14m_\phi^2}{3} & 0 \\ 0 & 2m_\phi^2
		\end{pmatrix}$ & $\begin{pmatrix}
		-2m_\phi^2 & 0\\0 & 2m_\phi^2\end{pmatrix}$ & $-2m_\phi^2$ & $2m_\phi^2$ & $2m_\phi^2$ & $2m_\phi^2$\\
		$C_{h_{3(5)}}$ & $\begin{pmatrix}
			-\frac{14}{3} & 0 & 0 \\
			0 & 2 & 0 \\
			0 & 0 & -2
		\end{pmatrix}$ & $\begin{pmatrix}
			-\frac{14}{3} & 0 \\
			0 & -2
		\end{pmatrix}$ & $\begin{pmatrix}
			2 & 0 \\ 0 & -2
		\end{pmatrix}$ & $2$ & $-2$ & $-2$ & $-2$
		\\\hline\hline
	\end{tabular}
\end{table*}
The LECs \(h_0\) and \(h_1\) can be determined by the masses of the \(D\) mesons. The spin average mass of the scalar charmed mesons and vector charmed mesons and their hyperfine splitting term at tree level read as~\cite{Gil-Dominguez:2023eld}
\begin{equation}\label{eq:average-splitting-charmed-meson}
	\begin{aligned}
		\frac{1}{4}\qty(M_P+3M_{P^*}) &= \frac{m_{\pi_0}^2 (a\delta_{1,2I} + 2\sigma)}{B_0} + 2m_s\qty(a\delta_{3l} + \sigma) + m_H,\\
		M_{P^*} - M_{P} &= \Delta + \frac{m_{\pi_0}^2 \qty(\delta_{1,2I} \Delta^{(a)} + 2\Delta^{(\sigma)})}{B_0} + 2m_s\qty(\delta_{3l} \Delta^{(a)} + \Delta^{(\sigma)}).
	\end{aligned}
\end{equation}
In Eq.~\eqref{eq:average-splitting-charmed-meson}, the $\delta_{1,2l}=1$, $\delta_{3l}=0$ for a $D^{(*)}$ meson, and $\delta_{1,2l}=0$, $\delta_{3l}=1$ for a $D_s^{(*)}$ meson. Substituting them into Eq.~\eqref{eq:average-splitting-charmed-meson}, we obtain,
\begin{equation}\label{eq:mD-mDs-tree}
	\begin{aligned}
		m_D &= \qty(m_H-\frac{3}{4}\qty(\Delta+2m_s\Delta^{(\sigma})+2m_s\sigma)+ \frac{\left( 4 \, a - 3 \, \Delta^{(a)} - 6 \, \Delta^{(\sigma)} + 8 \, \sigma \right)}{4 \, B_0}m_{\pi 0}^2 , \\
		m_{D_s} &= m_H+2am_s-\frac34\qty(\Delta+2m_s\qty(\Delta^{(a)}+\Delta^{(\sigma)}))+2m_s\sigma + \qty(\frac{2\sigma}{B_0}-\frac{3\Delta^{(\sigma)}}{2B_0})m_{\pi0}^2,
	\end{aligned}
\end{equation}
where  $\sigma=\frac{\sigma'}{m_\pi}B_0$, $a=\frac{a'm_\pi}{B_0}$, and $m_\pi$ is the physical pion mass. In Ref.~\cite{Liu:2012zya}, the masses of the $D$ and $D_s$ are give by
\begin{equation}\label{eq:mD-mDs-h01}
	\begin{aligned}
		m_D &= m_D^{(0)} + (h_1+2h_0) \frac{m_{\pi0}^2}{m_D^{(0)}},\\
		m_{D_s} & = m_{D_s}^{(0)} + 2h_0 \frac{m_{\pi0}^2}{m_{D_s}^{(0)}},
	\end{aligned}
\end{equation}
being $m_{D_{(s)}}^{(0)}$ the mass of the $D_{(s)}$ meson at leading order. By comparing the Eq.~\eqref{eq:mD-mDs-tree} and Eq.~\eqref{eq:mD-mDs-h01}, we obtain
\begin{equation}
	\begin{aligned}
		(h_1 + 2h_0)\frac{m_{\pi 0}^2}{m_D^{(0)}} &= \frac{\left( 4 \, a - 3 \, \Delta^{(a)} - 6 \, \Delta^{(\sigma)} + 8 \, \sigma \right)}{4 \, B_0}m_{\pi 0}^2, \\
		h_0 &= \frac{\qty(3\Delta^{(\sigma)}-4\sigma)\qty(-8am_s+3\Delta-4m_H-8\sigma m_s+6m_s\qty(\Delta^{(a)}+\Delta^{(\sigma)}))}{16B_0},
	\end{aligned}
\end{equation}
where $m_D^{(0)}=m_H - \frac34\qty(\Delta+2m_s\Delta^{(\sigma)})+2m_s\sigma$. Taking the values of $a$, $\Delta^{(a)}$, $\Delta^{(\sigma)}$, $\sigma$, and $m_H$ from Table~\ref{tab:Dpars}, the values of $h_0$ and $h_1$ read
\begin{equation}
	h_0 = 0.014,\, h_1=0.45.
\end{equation}
The value of $h_1$ is close to the one obtained in~\cite{Guo:2008gp} of $0.42$. 

Now we discuss the formalism used for the scattering amplitude to analyze the energy levels. The latter in the infinite volume reads
\begin{equation}
	T^{-1} = V ^{-1} - G,
\end{equation}
where \(G\) is a diagonal matrix with the {\it j}-th element \(G_j\) given by two-body propagator, i.e.,
\begin{equation}\label{eq:propagator}
	G_j(P) = i \int\frac{\dd^4{q}}{(2\pi)^4}\frac{1}{q^2 - m_j^2 + i\epsilon}\frac{1}{\qty(P-q)^2-M_j^2+i\epsilon}.
\end{equation}
\(m_j(M_j)\) are the hadron masses in a channel \(j\) and \(P\) the total four-momentum of two-meson system. The Eq.~\eqref{eq:propagator} can be evaluated using dimensional regularization (DR) method~\cite{Oller:1998zr} or a cutoff \(\Lambda\)~\cite{Oller:1998hw}. The subtraction constant \(a\) in DR scheme is given by~\cite{Zhuang:2024udv,Oller:2019opk}
\begin{equation}\label{eq:sub-cons}
	a(\Lambda) = -\frac{2}{(m_j+M_j)}\qty[m_j\log\qty(1+\sqrt{1+\frac{m_j^2}{\Lambda^2}}) + M_j\log\qty(1+\sqrt{1+\frac{M_j^2}{\Lambda^2}})] + 2\log\qty(\frac{\mu}{\Lambda}),
\end{equation}
The scale \(\mu\) is fixed at 1~GeV in this work and the UV cutoff \(\Lambda\) as a free parameter is determined by LQCD. Indeed, Note that Eq.~\eqref{eq:sub-cons} is independent on the scale \(\mu\)~\cite{Guo:2018tjx}. To better understand the subtraction constant dependence with the cutoff, we plot the \(a_{D\pi}\), \(a_{D\eta}\), and \(a_{D_s\bar{K}}\) in sector \(\qty(S, I)=\qty(0, 1/2)\) as a function of \(\Lambda\), see Fig.~\ref{fig:subtracion-Dpi}. In principle, a UV cutoff \(\Lambda\) larger than \(200\)~MeV~\cite{Albaladejo:2018mhb}, leads to values for the a subtraction constant of, \(a_i\lesssim -1.13\), taking the later more reasonable values around $-2$ for cutoffs in the range of $500-1000$~MeV.
\begin{figure*}[!htpb]
	\includegraphics[width=0.5\textwidth]{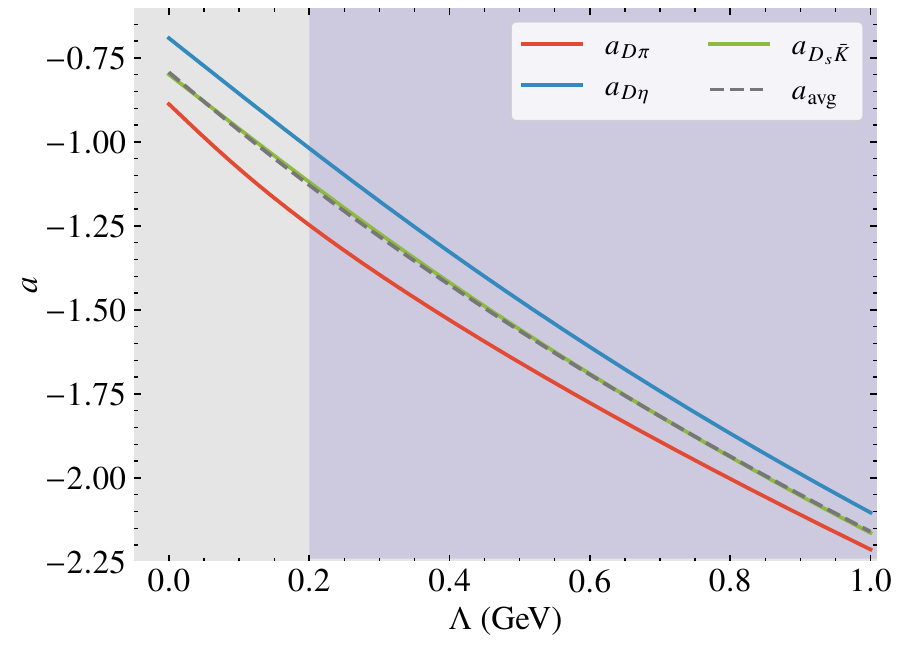}
	\caption{\label{fig:subtracion-Dpi} Cutoff dependence of the subtraction constants, \(a_{D\pi}\), \(a_{D\eta}\), \(a_{D_s\bar{K}}\), and their average value \(a_\mathrm{avg}\) from \(\Lambda=0\) to \(1\)~GeV.}
\end{figure*}

In the finite volume, the scattering amplitude is given by
\begin{equation}\label{eq:Tmat-finite}
	\tilde{T}^{-1} = V^{-1} - \tilde{G},
\end{equation}
where the \(\tilde{G}\) is a diagonal matrix in a finite volume. The element \(\tilde{G}_j\) in \(\tilde{G}\) can be evaluated in DR scheme, i.e.,
\begin{equation}
	\tilde{G}_j = G^\mathrm{DR}_j(P) + \lim_{q_\mathrm{max}\to\infty}\Delta \tilde{G}_j(P,q_\mathrm{max}),
\end{equation}
where the \(\Delta G_j\) reads~\cite{Doring:2011vk,Doring:2012eu,Doring:2013glu},
\begin{equation}\label{eq:diff-Gtilde}
	\Delta \tilde{G}_j = \frac{1}{L^3}\sum_{\vec{n}}^{q_\mathrm{max}}\frac{E}{P_0}I_j(\abs{\vec{q}^*(\vec{q})}) - G^\mathrm{cutoff}(P).
\end{equation}
In Eq.~\eqref{eq:diff-Gtilde}, the \(I(\abs{\vec{q}^*})\) is given by
\begin{equation}
	I(\abs{\vec{q}^{*}}) = \frac{1}{2\omega_{1j}\omega_{2j}}\frac{\omega_{1j}\omega_{2j}}{s-\qty(\omega_{1j}+\omega_{2j})^2+i\epsilon},
\end{equation}
where the \(\vec{q}^*\) is the momentum of two-body system at CM frame and \(\omega_{ij}^2=\abs{\vec{q}^*}^2+m_{ij}^2\). \(G^\mathrm{cutoff}\) is provided in, for example, ~\cite{Oller:1998hw}. In a moving frame with total four-momentum $P^\mu=(P^0,\vec{P})$, the \(\vec{q}^*\) is given by~\cite{Doring:2012eu}
\begin{equation}
	\vec{q}^*(\vec{q}) = \vec{q} + \qty[\qty(\frac{\sqrt{s}}{P^0}-1)\frac{\vec{q}\cdot\vec{P}}{\abs{\vec{P}}^2}-\frac{q^{*0}}{P^0}]\vec{P};\,\, 
	\vec{q} = \frac{2\pi}{L}\vec{n},\,\qty(\vec{n}\in\mathbb{Z}).
\end{equation}
In this work,  Eq.~\eqref{eq:diff-Gtilde} converges when $n=10$ and \(q_\mathrm{max}\simeq2000\)~MeV. The energy levels of a finite volume correspond to the poles in the Eq.~\eqref{eq:Tmat-finite}, i.e.,
\begin{equation}
	\det[\mathbb{I} - V\tilde{G}] = 0.
\end{equation}

\subsection{Heavy meson masses}
We reanalyze data for the low-lying charmed mesons from several lattice collaborations, in the same spirit that in Ref.~\cite{Gil-Dominguez:2023eld}. We use the framework of one-loop NLO HH$\chi$PT where the low energy constants are determined by analyzing the available lattice data from different LQCD simulations.

The one-loop masses of heavy mesons can be written in terms of two linear combinations, the spin average term, $\frac{1}{4}(M_{P_a}+3M_{P^*_a})$ and the hyperfine splitting, $M_{P^*}-M_P$, which respects and violate heavy quark spin symmetry, respectively. These, which have been calculated to one-loop NLO in Ref.~\cite{Jenkins:1992hx}, are
\begin{eqnarray}\label{eq:md1}
\frac{1}{4}(M_{P_l}+3M_{P^*_l})&= &\,m_H + \alpha_l-\sum_{X=\pi,K,\eta}\beta_l^{X}(g^2)\frac{M_X^3}{16\pi f_X^2} +\sum_{X=\pi,K,\eta}\left(\gamma_l^{X}-\lambda_l^{X}\alpha_l\right)\frac{M_X^2}{16\pi^2 f_X^2}\log\left(M_X^2/\mu^2\right)+c_l\\\label{eq:md2}
M_{P^*_l}-M_{P_l}&= &\,\Delta+\sum_{X=\pi,K,\eta}\left(\gamma_l^{X}-\lambda_l^{X}\Delta\right)\frac{M_X^2}{16\pi^2 f_X^2}\log\left(M_X^2/\mu^2\right)+\delta c_l\ 
\end{eqnarray}
where the coefficients $\alpha_l$, $\beta_l^{X}(g^2)$, $(\gamma_l^{X}-\lambda_l^{X}\alpha_l)$, $c_l$, $(\gamma_l^{X}-\lambda_l^{X}\Delta)$ and $\delta c_l$ are given in appendix A of Ref.~\cite{Gil-Dominguez:2023eld}. 

The coefficients $\alpha_l$, $(\gamma_l^{X}-\lambda_l^{X}\alpha_l)$, $c_l$, and $\delta c_l$ are proportional to powers of the light quark masses while $\beta_l^{X}(g^2)$ and $(\gamma_l^{X}-\lambda_l^{X}\Delta)$ accompany terms proportional to $M_X$, the mass of the $X$ pseudoscalar meson, $\pi$, $K$ or $\eta$. Thus, $m_H$ and $\Delta$ can be interpreted as the spin-average mass and hyperfine splitting of the heavy mesons in the chiral limit of SU(3). The scale $\mu$ in Eqs.~(\ref{eq:md1}) and (\ref{eq:md2}) is fixed to $770~$MeV. 

In the previous analysis from Ref.~\cite{Gil-Dominguez:2023eld}, the light meson mass dependence in cubic and logarithmic terms of Eqs.~(\ref{eq:md1}) and (\ref{eq:md2}) was driven by using the Gell-Mann-Okubo tree level mass formulas. However, in this work, we are using not only the light meson masses from the improved analysis of Ref~\cite{deconsnnlo} but also the explicit dependence on their decay constants, $f_\pi$,$f_K$ and $f_\eta$, extracted from the same work.
Another difference is the addition of data from Ref.\cite{Yeo:2024chk} at the SU(3) flavor symmetric point for both; obtaining information about this flavor symmetric limit and to better constrain the fit at high pion masses.

By doing so, we extend our analysis up to a mass of $\sim 700$ MeV. In order to do that we needed to fix the axial vector coupling constant that enters in $\beta_l^{X}(g^2)$ coefficients to $g^2=0.32$. In that way, we can have a correct behavior for the charmed meson masses in the high pion mass region. The best values for the parameters of the analysis are shown in Table \ref{tab:Dpars}. The results for the global fit to the charmed meson masses in lattice units are shown in Fig.~\ref{fig:Dmasseslat}. We see a very nice description of all the data, with a $\chi^2/\mathrm{d.o.f} = 0.88$. Our final results for the pion mass dependence of the $D$ and $D_s$ meson masses for physical charm and strange quark mass are shown in Fig.~\ref{fig:mDmDs}.

\begin{figure*}[!htpb]
    \centering
	\includegraphics[width=0.9\columnwidth]{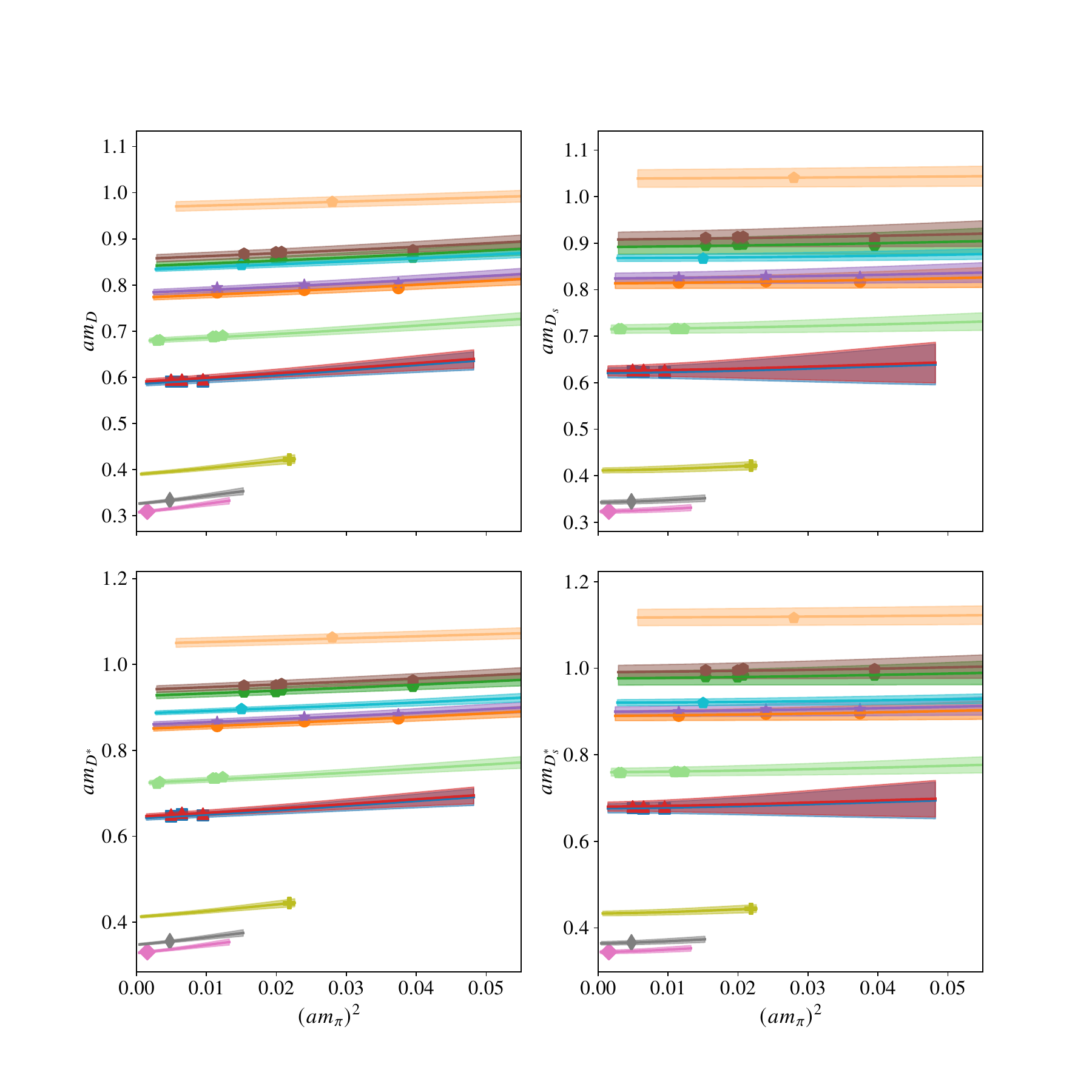}
    \resizebox{0.9\columnwidth}{!}{\input{figure/fig_amDmesons_legend.pgf}}
	\caption{\label{fig:Dmasseslat} Results for the fit to the charmed meson masses, in lattice units (solid lines with error bars). Lattice errors are not displayed for clarity, being comparable or smaller than the marker size.}
\end{figure*}
\begin{figure*}[!htpb]
	\includegraphics[width=0.5\textwidth]{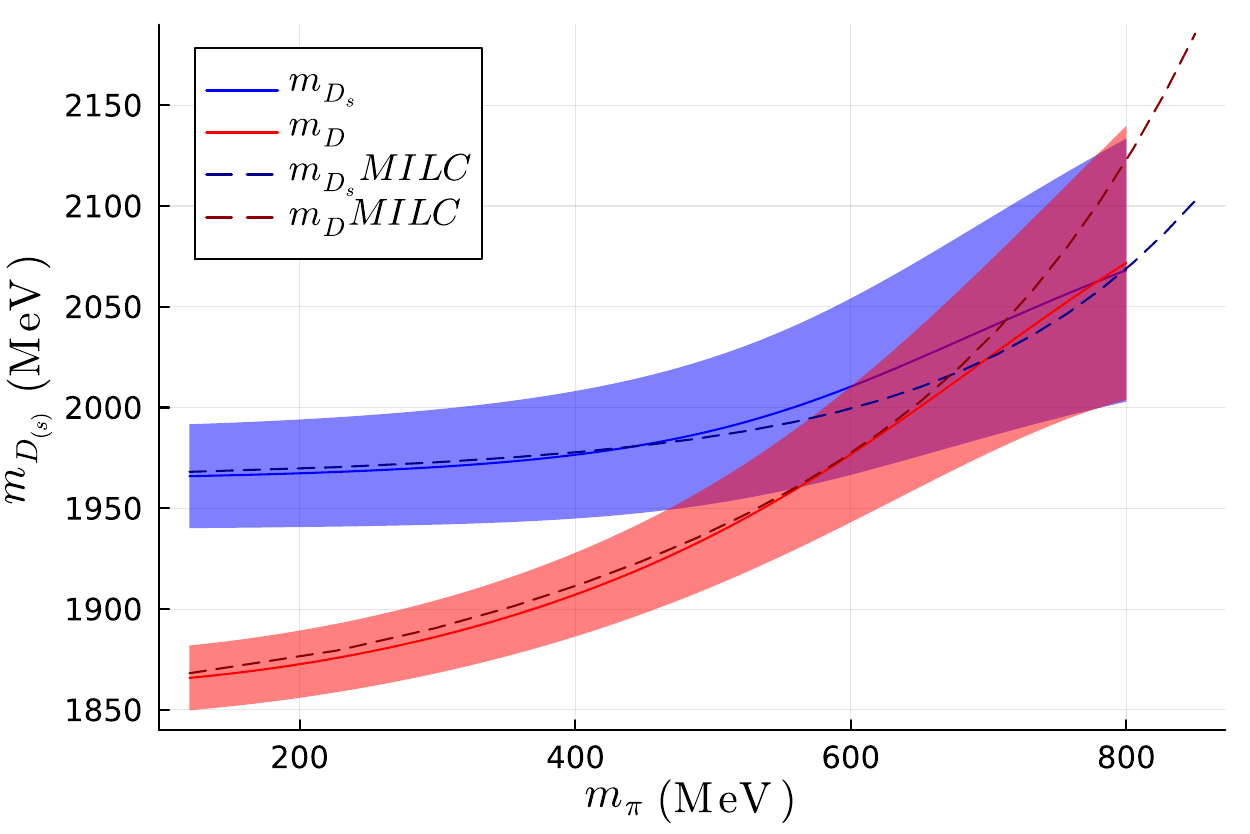}
	\caption{\label{fig:mDmDs} Pion mass dependence of the $D$ and $D_s$ meson masses (in MeV). The charmed quark mass is fixed to the physical value. Solid lines correspond to our results, while dashed lines represent the extrapolation performed by the MILC collaboration \cite{Brambilla:2017hcq}. The error band represents the uncertainty arising from the errors of the parameters obtained from the fit.}
\end{figure*}

\begin{table}[h!]
\renewcommand{\arraystretch}{2.5}
	\setlength{\tabcolsep}{0.45cm}
	\centering
\caption{\label{tab:Dpars}Collection of parameters from the fit to charmed meson masses.}
\begin{tabular}{ccccccc} \hline\hline
$\frac{\sigma m_\pi}{{B_0}}\cdot 10^{2}$ & $\frac{a m_\pi}{{B_0}}\cdot 10^{2}$ & $\frac{b m^3_\pi}{{B_0}^{2}}\cdot 10^{4}$& $\frac{c m^3_\pi}{{B_0}^{2}}\cdot 10^{4}$ & $\frac{d m^3_\pi}{{B_0}^{2}}\cdot 10^{4}$ &  $\frac{\Delta^{(\sigma)}m_\pi}{{B_0}}\cdot 10^{4}$& $\frac{\Delta^{(a)}m_\pi}{{B_0}}\cdot 10^{4}$\\ \hline

$0.15 \pm 0.20$ & $3.31 \pm 0.18$ & $-2.99 \pm 1.84$ & $2.99\pm 1.05$ & $2.87\pm 0.34$ & $7.14\pm 3.58$ & $-8.13\pm 3.01$ \\ \hline\hline
\end{tabular}
\end{table}

\subsection{Lattice setup and some results}
In Table~\ref{tab:hadspe_setting}, we summarize the lattice setups used for analyzing the energy levels. In Fig.~\ref{fig:ecm-lo-nlo}, we reproduce the energy levels in different sectors by the free parameters fixed by the energy levels in LQCD simulations~\cite{Moir:2016srx,Gayer:2021xzv,Cheung:2020mql,Yeo:2024chk}. The corresponding correction matrix between the parameters is given Fig.~\ref{fig:correlation}, which are non-negligible. The poles in different sectors evaluated in this work are given in Table~\ref{tab:pole-lo}. In addition, we make the predictions for the scattering lengths and the phase shifts shown in Fig.~\ref{fig:sl-ps-nlo}. By comparison with results in~\cite{Gayer:2021xzv,Cheung:2020mql,Moir:2016srx,Yeo:2024chk,Liu:2012zya,Mohler:2013rwa,Lang:2014yfa,Yan:2024yuq}, we observe that our model can explain the scattering lengths and the phase shifts well. We also predict the invariant mass spectrum in the \(D\pi\) final state at the physical point in Fig.~\ref{fig:dist-coupling}~(Left). As we can see, while the $D_0^*(2100)$ is clearly seen as a peak, the higher pole appears as a bump close to the $D\eta$ and $D_s\bar{K}$ thresholds. In Fig.~\ref{fig:dist-coupling}~(Right), we show the pion mass dependence of the couplings between the \(D_0^*(2100)\) and channels \(D\pi\), \(D\eta\), and \(D_s\bar{K}\). The derivatives of the couplings are discontinuous at \(m_\pi=310\), and \(420\)~MeV, when the lower-energy pole becomes virtual and bound, respectively. Finally, the pion mass dependence of the poles in different sectors are presented in Fig.~\ref{fig:traj-pole-SI}.
\begin{table*}[!hbtp]
	\renewcommand{\arraystretch}{1.3}
	\setlength{\tabcolsep}{0.16cm}
	\centering
	\caption{\label{tab:pole-lo}The poles for $(S,I)=(0,1/2),\,(1,0),\,(-1,0),\,(1,1)$ at LO and up to NLO, respectively. The dimensions of the pole positions are MeV, while the units of the couplings are GeV. The $-$ and $+$ present the unphysical sheet and the physical sheet, respectively.}
	\begin{tabular}{lcccccc}
		\hline\hline
		$m_\pi$~(MeV) & \multicolumn{2}{c}{$138$} & \multicolumn{2}{c}{$239$} & \multicolumn{2}{c}{$391$} \\\hline
		\multicolumn{7}{c}{LO}\\\cline{2-7}
		\multicolumn{7}{c}{\((0, 1/2)\)}\\
		Threshold / MeV & \multicolumn{2}{c}{$\{2005,2413,2464\}$} & \multicolumn{2}{c}{$\{2119,2445,2474\}$} & \multicolumn{2}{c}{$\{2277,2473,2501\}$}\\
		R.S. & $[-++]$ & $[---]$ & $[-++]$ & $[---]$ & $[+++]$ & $[---]$\\
		Pole & $2089(6)-i85(13)$ & $2303(6)-i150(20)$ & $2132(7)-i65(17)$ & $2329(5)-i169(21)$ & $2276(2)$ & $2369(2)-i197(22)$ \\
		$\abs{g_{D\pi}}$ & $9.0(3)$ & $7.6(1)$ & $10.0(1)$ & $7.8(1)$ & $2.9(14)$ & $8.0(1)$\\
		$\abs{g_{D\eta}}$ & $0.1(1)$ & $8.2(1)$ & $0.1(1)$ & $7.8(1)$ & $0.1(1)$ & $7.7(1)$\\
		$\abs{g_{D_s\bar{K}}}$ & $3.7(2)$ & $7.6(1)$ & $4.5(1)$ & $7.1(1)$ & $1.6(8)$ & $6.5(1)$\\
		\hline
		\multicolumn{7}{c}{\((1,0)\)} \\
		Threshold / MeV & \multicolumn{2}{c}{$\{2362,2516\}$} & \multicolumn{2}{c}{$\{2387,2532\}$} & \multicolumn{2}{c}{$\{2435,2538\}$}\\
		R.S. & \multicolumn{2}{c}{$[++]$} & \multicolumn{2}{c}{$[++]$} & \multicolumn{2}{c}{$[++]$}\\
		LQCD & \multicolumn{2}{c}{$\cdots$}& \multicolumn{2}{c}{$2362(3)$} & \multicolumn{2}{c}{$2380(3)$} \\
		Pole & \multicolumn{2}{c}{$2348(12)$} & \multicolumn{2}{c}{$2375(12)$} & \multicolumn{2}{c}{$2421(12)$} \\
		$\abs{g_{DK}}$ & \multicolumn{2}{c}{$7.8(17)$} & \multicolumn{2}{c}{$7.7(20)$} & \multicolumn{2}{c}{$7.9(18)$}\\
		$\abs{g_{D_s\eta}}$ & \multicolumn{2}{c}{$4.9(10)$} & \multicolumn{2}{c}{$4.9(10)$} & \multicolumn{2}{c}{$5.0(10)$}\\\hline
		\multicolumn{7}{c}{\((-1,0)\)} \\
		Threshold / MeV & \multicolumn{2}{c}{$\{2362\}$} & \multicolumn{2}{c}{$\{2387\}$} & \multicolumn{2}{c}{$\{2435\}$}\\
		R.S. & \multicolumn{2}{c}{$[-]$} & \multicolumn{2}{c}{$[-]$} & \multicolumn{2}{c}{$[-]$}\\
		LQCD & \multicolumn{2}{c}{$\cdots$} & \multicolumn{2}{c}{$2170(140)$} & \multicolumn{2}{c}{$2310(90)$}\\
		Pole & \multicolumn{2}{c}{$2316(1)-i166(24)$} & \multicolumn{2}{c}{$2342(1)-i180(23)$} & \multicolumn{2}{c}{$2376(1)-i196(24)$}\\
		$\abs{g_{D\bar{K}}}$ & \multicolumn{2}{c}{$12.5(2)$} & \multicolumn{2}{c}{$12.5(2)$} & \multicolumn{2}{c}{$12.8(2)$}\\\hline
		\multicolumn{7}{c}{\((1,1)\)} \\
		Threshold / MeV & \multicolumn{2}{c}{$\{2107, 2362\}$} & \multicolumn{2}{c}{$\cdots$} & \multicolumn{2}{c}{$\cdots$}\\
		R.S. & \multicolumn{2}{c}{$[--]$} & \multicolumn{2}{c}{$\cdots$} & \multicolumn{2}{c}{$\cdots$}\\
		Pole & \multicolumn{2}{c}{$2299(3)-i182(18)$}& \multicolumn{2}{c}{$\cdots$} & \multicolumn{2}{c}{$\cdots$}\\
		$\abs{g_{D_s\pi}}$ & \multicolumn{2}{c}{$8.5(1)$} & \multicolumn{2}{c}{$\cdots$} & \multicolumn{2}{c}{$\cdots$} \\
		$\abs{g_{DK}}$ & \multicolumn{2}{c}{$7.5(1)$} & \multicolumn{2}{c}{$\cdots$} & \multicolumn{2}{c}{$\cdots$}\\\hline\hline
		\multicolumn{7}{c}{Up to NLO}\\\cline{2-7}
		\multicolumn{7}{c}{\((1,0)\)}\\
		R.S. & \multicolumn{2}{c}{$[++]$} & \multicolumn{2}{c}{$[++]$} & \multicolumn{2}{c}{$[++]$}\\
		Pole & \multicolumn{2}{c}{$2333(18)$} & \multicolumn{2}{c}{$2362(17)$} & \multicolumn{2}{c}{$2406(18)$} \\
		$\abs{g_{DK}}$ & \multicolumn{2}{c}{$10.3(16)$} & \multicolumn{2}{c}{$10.0(17)$} & \multicolumn{2}{c}{$10.3(18)$}\\
		$\abs{g_{D_s\eta}}$ & \multicolumn{2}{c}{$6.3(10)$} & \multicolumn{2}{c}{$6.3(11)$} & \multicolumn{2}{c}{$6.5(11)$}\\\hline
		\multicolumn{7}{c}{\((-1,0)\)} \\
		R.S. & \multicolumn{2}{c}{$[-]$} & \multicolumn{2}{c}{$[-]$} & \multicolumn{2}{c}{$[-]$}\\
		
		Pole & \multicolumn{2}{c}{$2506_{(42)(266)}^{(60)(2)}-i126_{(63)(63)}^{(60)(7)}$} & \multicolumn{2}{c}{$2526_{(42)(264)}^{(180)(112)}-i126_{(63)(63)}^{(62)(7)}$} & \multicolumn{2}{c}{$2574_{(42)(287)}^{(180)(112)}-i126_{(63)(63)}^{(61)(7)}$}\\
		$\abs{g_{D\bar{K}}}$ & \multicolumn{2}{c}{$8.7(6)$} & \multicolumn{2}{c}{$8.8(6)$} & \multicolumn{2}{c}{$8.8(6)$}\\\hline
		\multicolumn{7}{c}{\((1,1)\)}\\
		R.S. & \multicolumn{2}{c}{$[--]$} & \multicolumn{2}{c}{$\cdots$} & \multicolumn{2}{c}{$\cdots$}\\
		Pole & \multicolumn{2}{c}{$2398(41)-i123(12)$}& \multicolumn{2}{c}{$\cdots$} & \multicolumn{2}{c}{$\cdots$}\\
		$\abs{g_{D_s\pi}}$ & \multicolumn{2}{c}{$6.9(2)$} & \multicolumn{2}{c}{$\cdots$} & \multicolumn{2}{c}{$\cdots$} \\
		$\abs{g_{DK}}$ & \multicolumn{2}{c}{$6.1(1)$} & \multicolumn{2}{c}{$\cdots$} & \multicolumn{2}{c}{$\cdots$}
		\\\hline\hline
	\end{tabular}
\end{table*}
\begin{table*}[!htbp]
	\centering
	\caption{\label{tab:hadspe_setting}A summary of lattices setups for $m_\pi=239, 284, 327, 391$ MeV~\cite{Moir:2016srx,Gayer:2021xzv,Cheung:2020mql,Wilson:2019wfr}, and $688$ MeV~\cite{Yeo:2024chk,Edwards:2012fx,Woss:2018irj}. The errors of the $a_s$ and $a_t$ are propagated from the $a_s=a_t\xi$ and $a_t^{-1}=\frac{m_{\Omega,\mathrm{phy}}}{a_tm_\Omega}$. The anisotropy used in the SU(3) favor symmetry is the mean of $\xi_{\eta_8}=3.446(3)$~\cite{Edwards:2012fx}, $\xi_{D_{\bar{3}}}=3.468(2)$, $\xi_{D_{\bar{3}}^*|_{\lambda=0}}=3.492(4)$, $\xi_{D_{\bar{3}}^*|_{|\lambda|=1}}=3.476(4)$~\cite{Yeo:2024chk}.}
	\renewcommand{\arraystretch}{2.5}
	\setlength{\tabcolsep}{0.15cm}
	\begin{tabular}{lcccccccc}\hline\hline
		$\qty(L/a_s)^3\times T/a_t$ & $m_\pi/\mathrm{MeV}$ & $a_t^{-1}/\mathrm{MeV}$ & $a_tm_\pi$ & $a_tm_K$ & $a_tm_\eta$ & $a_tm_\Omega$ & $\xi$ & $a_s/\mathrm{fm}$\\\hline
		$32^3\times256$ & $239$ & $6079(13)$ & $0.03928(18)$ & $0.08344(7)$ & $0.09299(56)$  & $0.2751(6)$ & $3.453(6)$ & $0.1121(3)$\\ 
		$24^3\times256$ & $284$ & $5988(17)$ & $0.04735(22)$ & $0.08659(14)$ & $0.09602(70)$ & $0.2793(8)$ & $3.455(6)$ & $0.1139(4)$\\ 
		$24^3\times256$ & $327$ & $5854(16)$ & $0.05593(28)$ & $0.09027(15)$ & $0.09790(100)$ & $0.2857(8)$ & $3.456(9)$ & $0.1165(4)$\\ 
		$\qty{16^3,20^3,24^3}\times128$ & $391$ & $5667(42)$ & $0.06906(13)$ & $0.09698(9)$ & $0.10364(19)$ & $0.2951(22)$ & $3.444(6)$ & $0.1199(9)$\\ 
		$\qty{16^3,20^3,24^3}\times128$ & $688$ & $4655(9)$ & $0.1478(1)$ & $0.1478(1)$ & $0.1478(1)$ & $0.3593(7)$ & $3.471(28)$ & $0.1471(12)$\\\hline\hline
	\end{tabular}\\[0.5cm]
	\begin{tabular}{cccccc}\hline\hline
		$m_\pi$ & $m_K$ & $m_\eta$ & $m_D$ & $m_{D_s}$ & $m_{\Omega}$ \\\hline
		$239(1)$ & $507(1)$ & $565(3)$ & $1880(1)$ & $1967(1)$ & $1672(4)$ \\
		$391(1)$ & $550(1)$ & $587(1)$ & $1885(1)$ & $1951(1)$ & $1672(12)$ \\
		$688(1)$ & $\cdots$ & $\cdots$ & $\cdots$ & $1962(1)$ & $1672(3)$\\\hline\hline
	\end{tabular}
\end{table*}

\begin{figure}[!ht]
	\centering
	\includegraphics[width=0.5\textwidth]{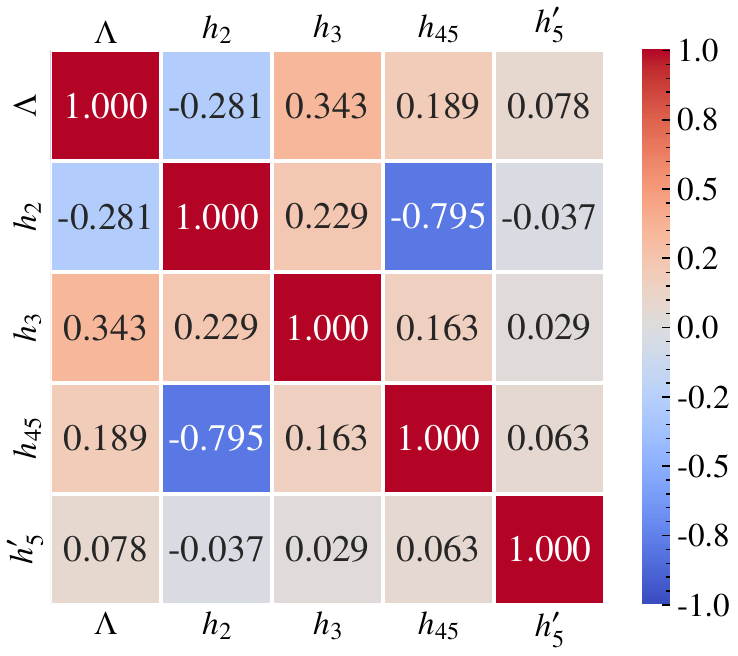}
	\caption{\label{fig:correlation}The correlations between the parameters. The reduced-$\chi^2$ obtained in this work is $\chi^2/\mathrm{d.o.f}=1.5$.}
\end{figure}

\begin{figure*}[!htpb]
	\centering
	\includegraphics[width=0.3\textwidth]{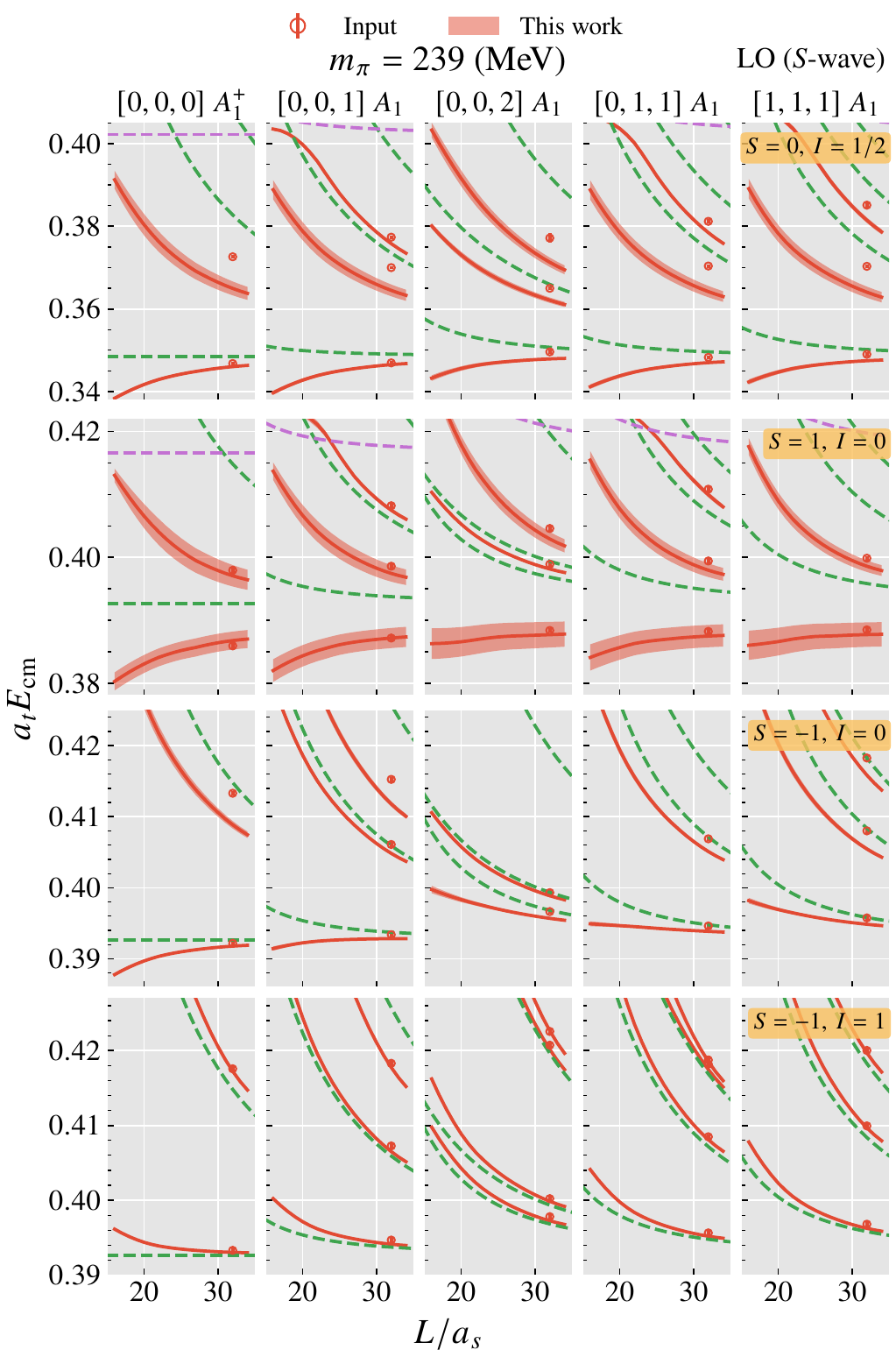}
	\hspace{3pt}
	\includegraphics[width=0.3\textwidth]{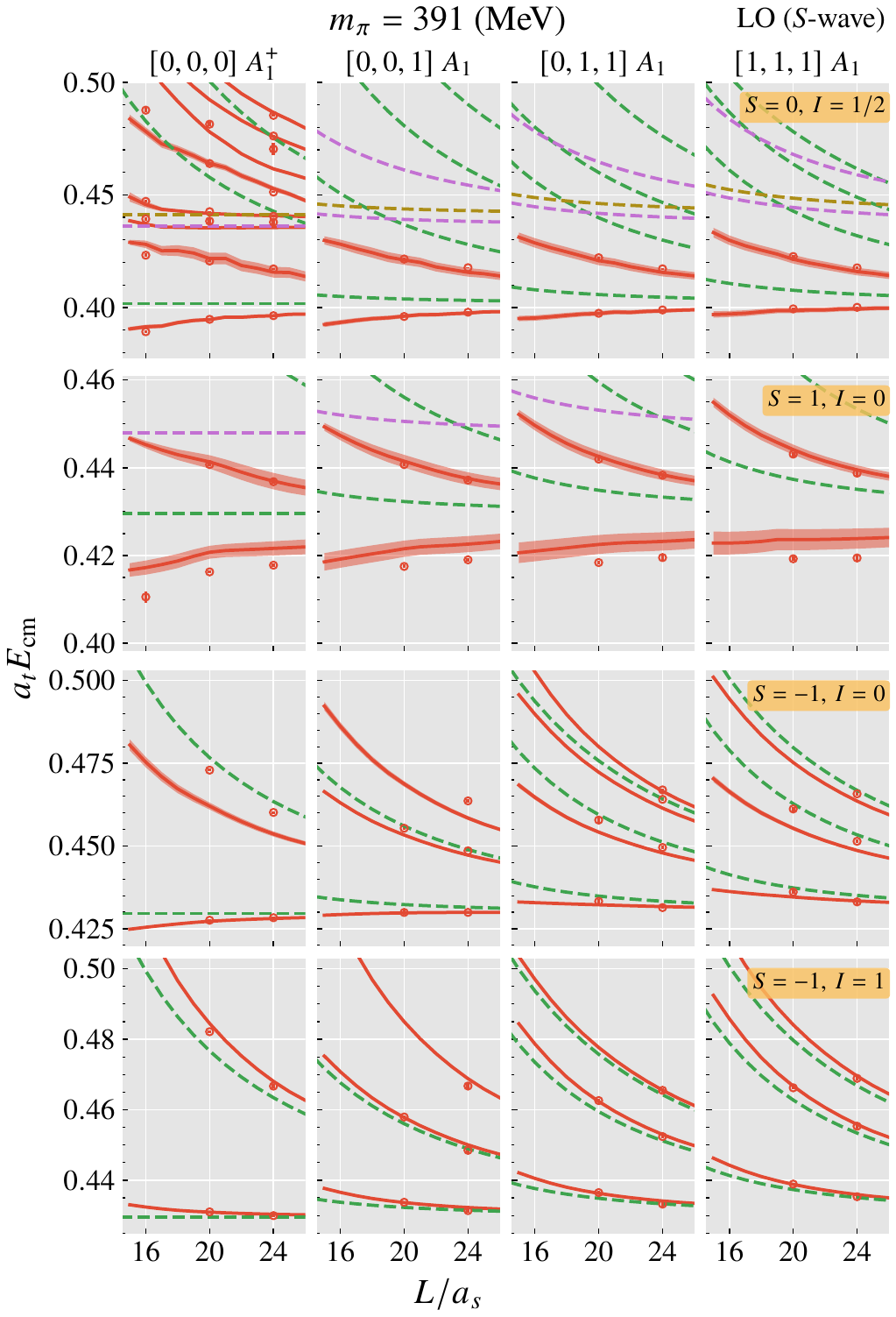}
	\hspace{3pt}
	\includegraphics[width=0.3\textwidth]{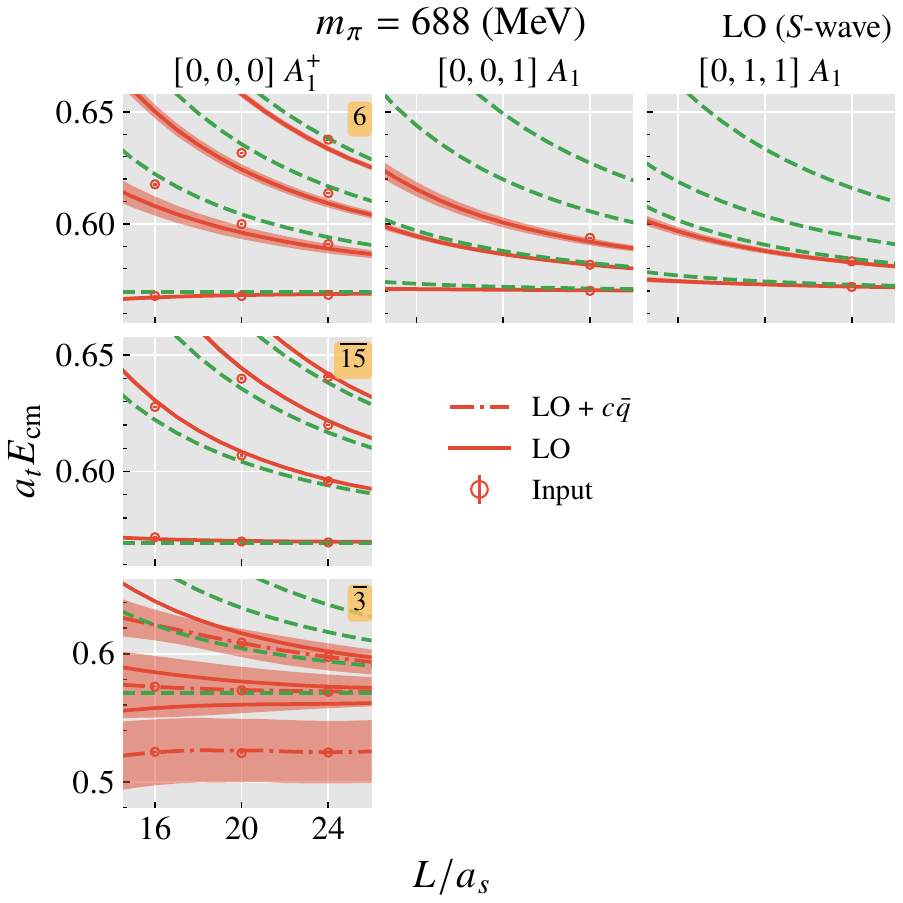}
	\noindent\rule{0.9\textwidth}{0.4pt}
	\includegraphics[width=0.3\textwidth]{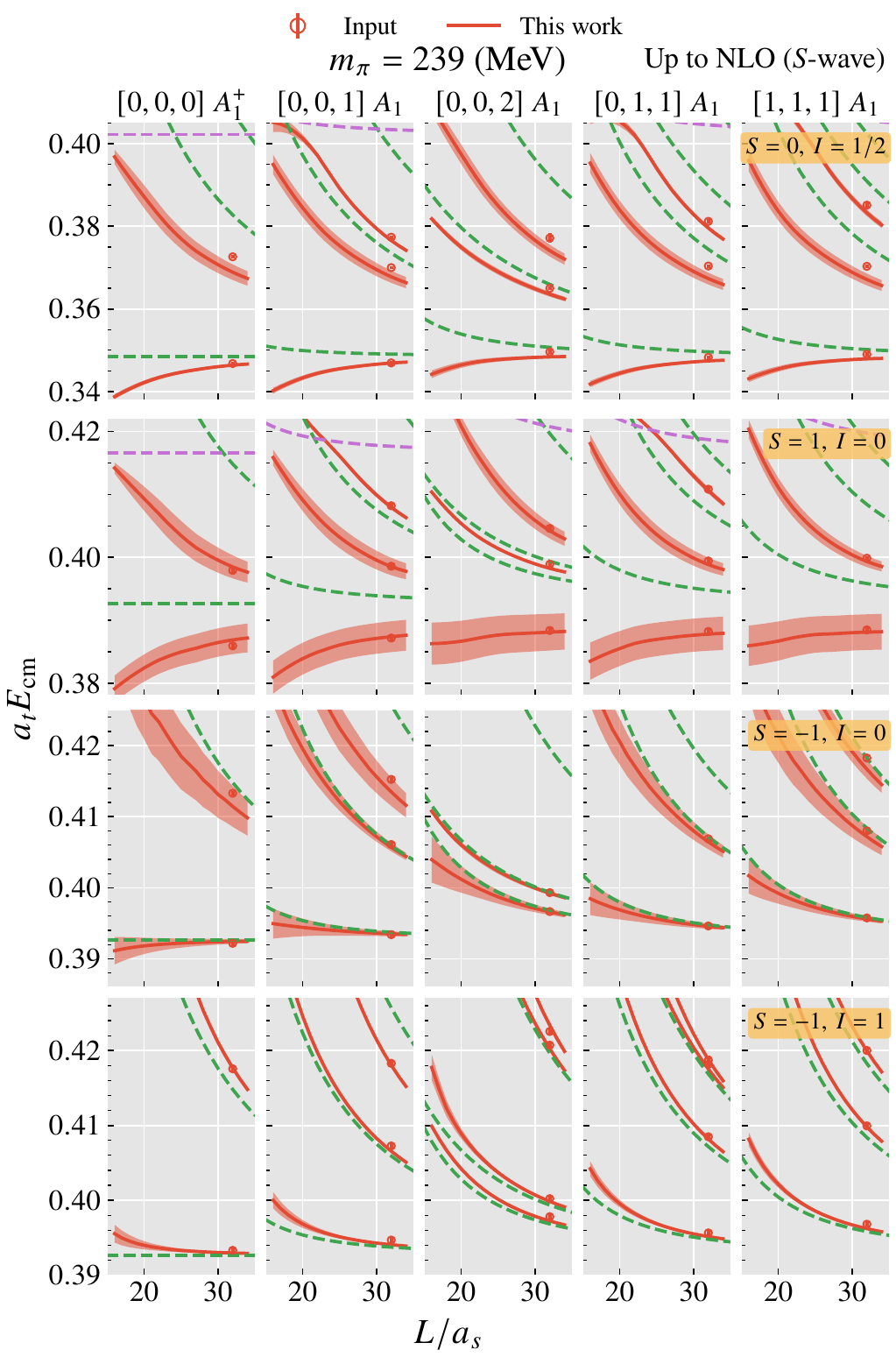}
	\hspace{3pt}
	\includegraphics[width=0.3\textwidth]{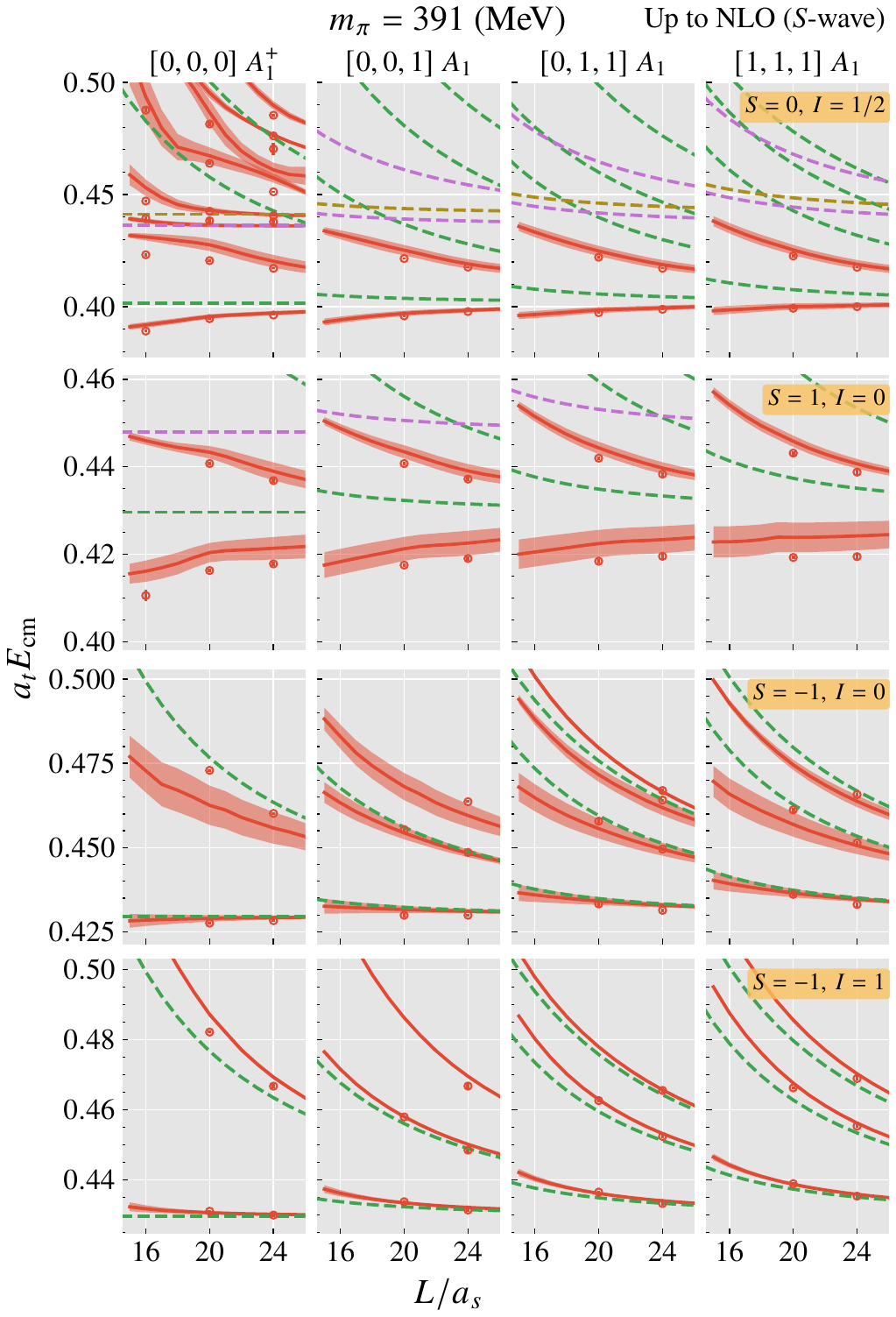}
	\hspace{3pt}
	\includegraphics[width=0.3\textwidth]{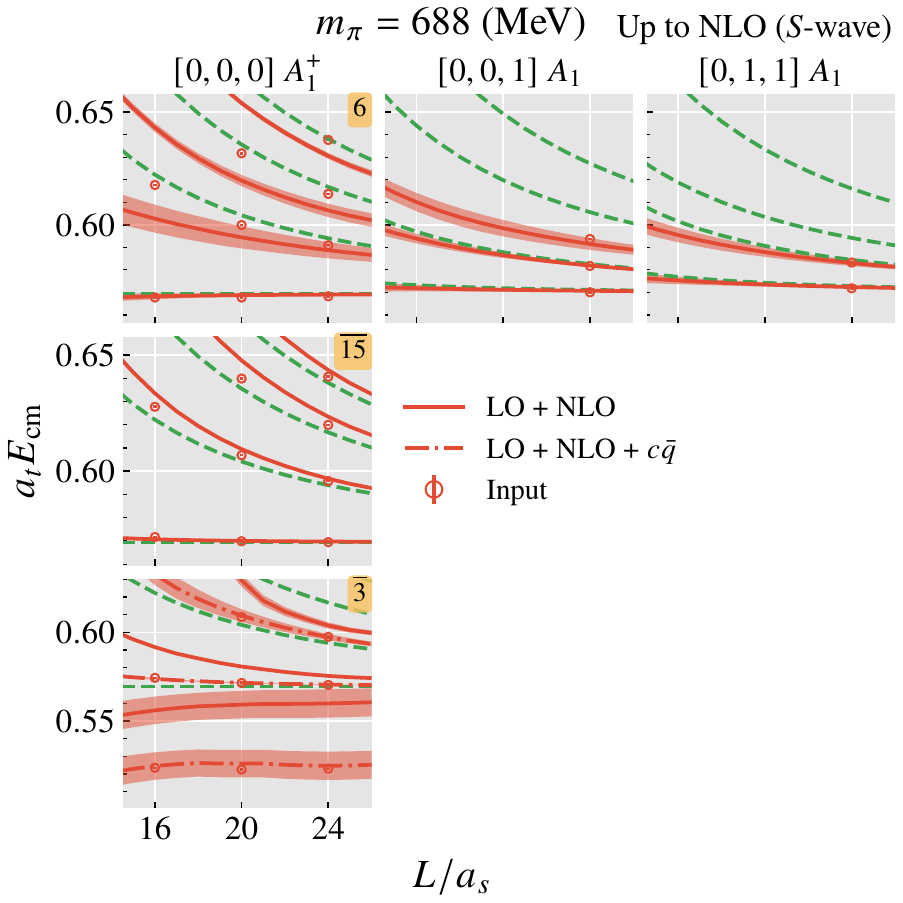}
	\caption{\label{fig:ecm-lo-nlo}The result from the UChPT fits in comparison to the energy levels~\cite{Moir:2016srx,Gayer:2021xzv,Cheung:2020mql,Yeo:2024chk} as input to constrain the free parameters at LO and up to NLO. The dashed lines denote the free energy levels.}
\end{figure*}

\begin{figure*}
	\centering
	\resizebox{0.8\textwidth}{!}{\input{figure/scattering_length_legend.pgf}}
	\includegraphics[width=0.9\textwidth]{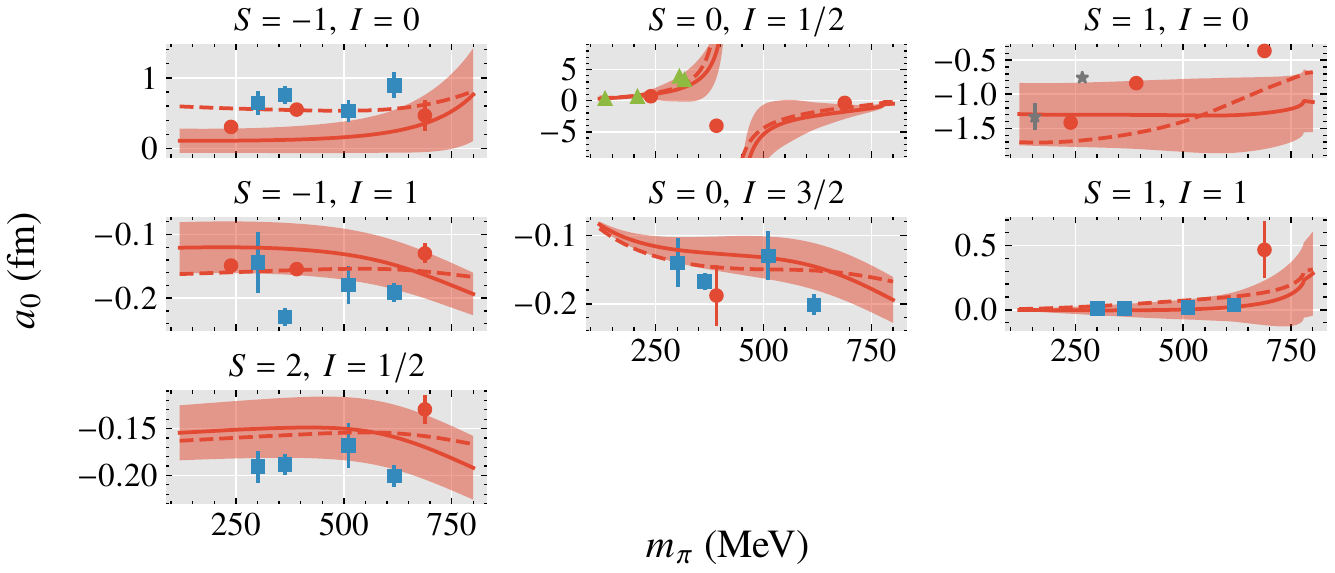}
	\vspace{4pt}
	\noindent\rule{0.9\textwidth}{0.4pt}
	\vspace{4pt}
		\resizebox{0.9\textwidth}{!}{
\begingroup%
\makeatletter%
\begin{pgfpicture}%
\pgfpathrectangle{\pgfpointorigin}{\pgfqpoint{6.300000in}{0.408000in}}%
\pgfusepath{use as bounding box, clip}%
\begin{pgfscope}%
\pgfsetbuttcap%
\pgfsetmiterjoin%
\pgfsetlinewidth{0.000000pt}%
\definecolor{currentstroke}{rgb}{0.500000,0.500000,0.500000}%
\pgfsetstrokecolor{currentstroke}%
\pgfsetdash{}{0pt}%
\pgfpathmoveto{\pgfqpoint{0.000000in}{0.000000in}}%
\pgfpathlineto{\pgfqpoint{6.300000in}{0.000000in}}%
\pgfpathlineto{\pgfqpoint{6.300000in}{0.408000in}}%
\pgfpathlineto{\pgfqpoint{0.000000in}{0.408000in}}%
\pgfpathlineto{\pgfqpoint{0.000000in}{0.000000in}}%
\pgfpathclose%
\pgfusepath{}%
\end{pgfscope}%
\begin{pgfscope}%
\pgfsetbuttcap%
\pgfsetmiterjoin%
\pgfsetlinewidth{0.000000pt}%
\definecolor{currentstroke}{rgb}{0.000000,0.000000,0.000000}%
\pgfsetstrokecolor{currentstroke}%
\pgfsetstrokeopacity{0.000000}%
\pgfsetdash{}{0pt}%
\pgfpathmoveto{\pgfqpoint{0.050000in}{0.050000in}}%
\pgfpathlineto{\pgfqpoint{6.250000in}{0.050000in}}%
\pgfpathlineto{\pgfqpoint{6.250000in}{0.358000in}}%
\pgfpathlineto{\pgfqpoint{0.050000in}{0.358000in}}%
\pgfpathlineto{\pgfqpoint{0.050000in}{0.050000in}}%
\pgfpathclose%
\pgfusepath{}%
\end{pgfscope}%
\begin{pgfscope}%
\pgfsetbuttcap%
\pgfsetroundjoin%
\pgfsetlinewidth{1.806750pt}%
\definecolor{currentstroke}{rgb}{0.886275,0.290196,0.200000}%
\pgfsetstrokecolor{currentstroke}%
\pgfsetdash{}{0pt}%
\pgfpathmoveto{\pgfqpoint{4.472333in}{0.140708in}}%
\pgfpathlineto{\pgfqpoint{4.472333in}{0.307375in}}%
\pgfusepath{stroke}%
\end{pgfscope}%
\begin{pgfscope}%
\pgfsetbuttcap%
\pgfsetmiterjoin%
\definecolor{currentfill}{rgb}{0.886275,0.290196,0.200000}%
\pgfsetfillcolor{currentfill}%
\pgfsetlinewidth{1.003750pt}%
\definecolor{currentstroke}{rgb}{0.886275,0.290196,0.200000}%
\pgfsetstrokecolor{currentstroke}%
\pgfsetdash{}{0pt}%
\pgfsys@defobject{currentmarker}{\pgfqpoint{-0.041667in}{-0.041667in}}{\pgfqpoint{0.041667in}{0.041667in}}{%
\pgfpathmoveto{\pgfqpoint{-0.041667in}{-0.041667in}}%
\pgfpathlineto{\pgfqpoint{0.041667in}{-0.041667in}}%
\pgfpathlineto{\pgfqpoint{0.041667in}{0.041667in}}%
\pgfpathlineto{\pgfqpoint{-0.041667in}{0.041667in}}%
\pgfpathlineto{\pgfqpoint{-0.041667in}{-0.041667in}}%
\pgfpathclose%
\pgfusepath{stroke,fill}%
}%
\begin{pgfscope}%
\pgfsys@transformshift{4.472333in}{0.224042in}%
\pgfsys@useobject{currentmarker}{}%
\end{pgfscope}%
\end{pgfscope}%
\begin{pgfscope}%
\definecolor{textcolor}{rgb}{0.000000,0.000000,0.000000}%
\pgfsetstrokecolor{textcolor}%
\pgfsetfillcolor{textcolor}%
\pgftext[x=4.772333in,y=0.165708in,left,base]{\color{textcolor}{\rmfamily\fontsize{12.000000}{14.400000}\selectfont\catcode`\^=\active\def^{\ifmmode\sp\else\^{}\fi}\catcode`\%=\active\def
\end{pgfscope}%
\end{pgfpicture}%
\makeatother%
\endgroup%
}
	\includegraphics[width=0.9\textwidth]{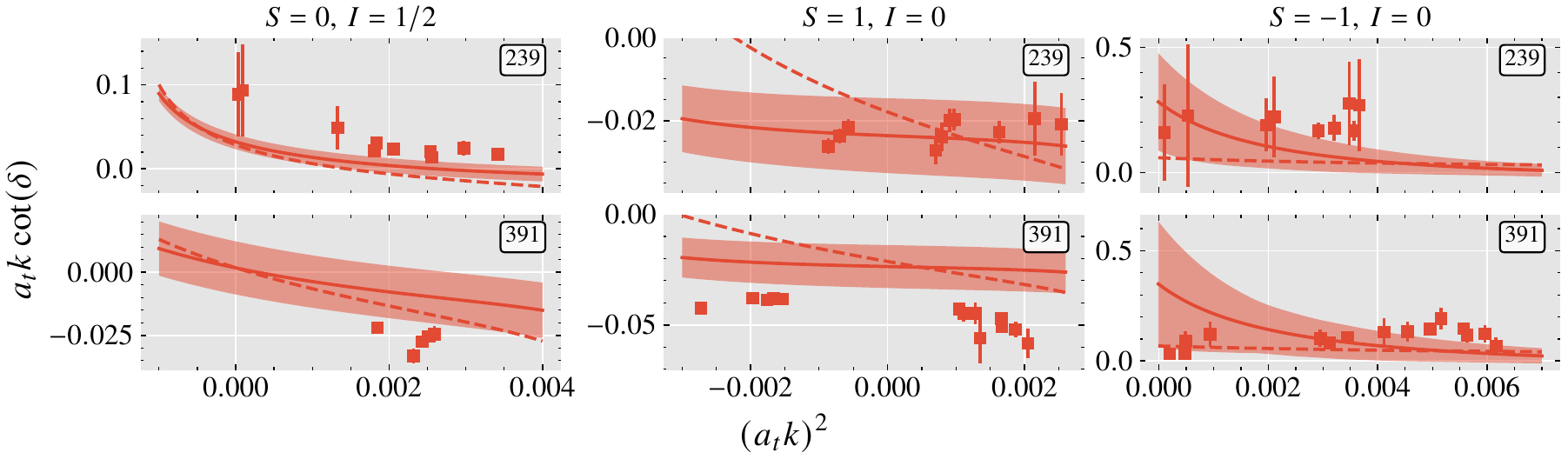}
	\noindent\rule{0.9\textwidth}{0.4pt}
	\includegraphics[width=0.9\textwidth]{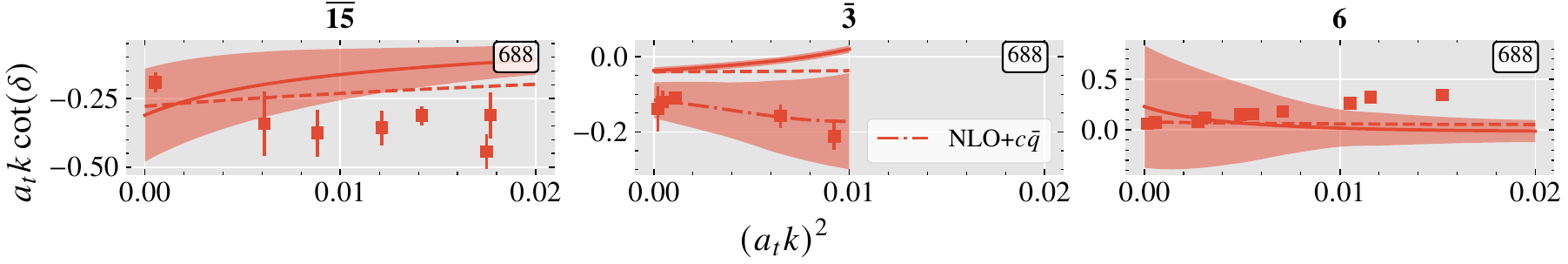}
	\caption{\label{fig:sl-ps-nlo}{\it S}-wave scattering lengths and phase shifts predicted by this work and compared with Refs.~\cite{Gayer:2021xzv,Cheung:2020mql,Moir:2016srx,Yeo:2024chk,Liu:2012zya,Mohler:2013rwa,Lang:2014yfa,Yan:2024yuq}.}
\end{figure*}

\begin{figure*}[htbp]
    \centering
    \includegraphics[width=0.48\textwidth]{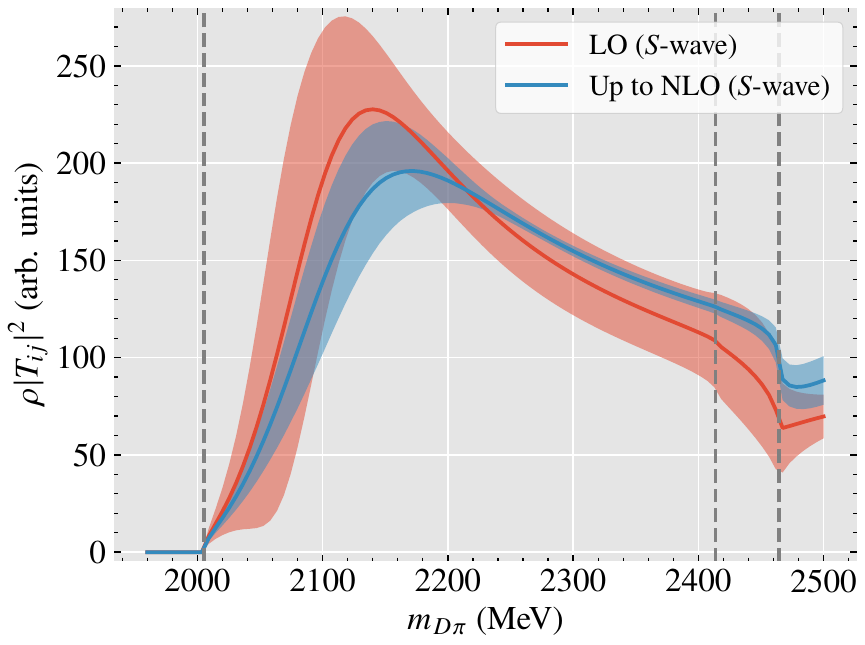}
    \hfill
    \includegraphics[width=0.48\textwidth]{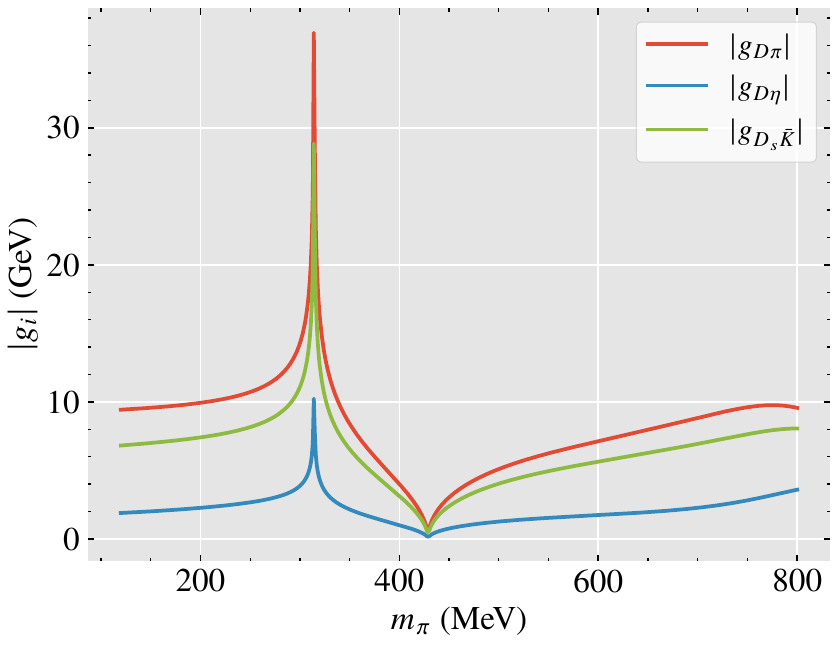}
    \caption{\label{fig:dist-coupling}Left: Predictions for the invariant mass spectrum in the $D\pi$ final state. Right: The pion mass dependence of the couplings between the $D_0^*(2100)$ and the channels $D\pi$, $D\eta$, and $D_s\bar{K}$.}
\end{figure*}

\begin{figure*}
	\centering
	\includegraphics[width=0.9\textwidth]{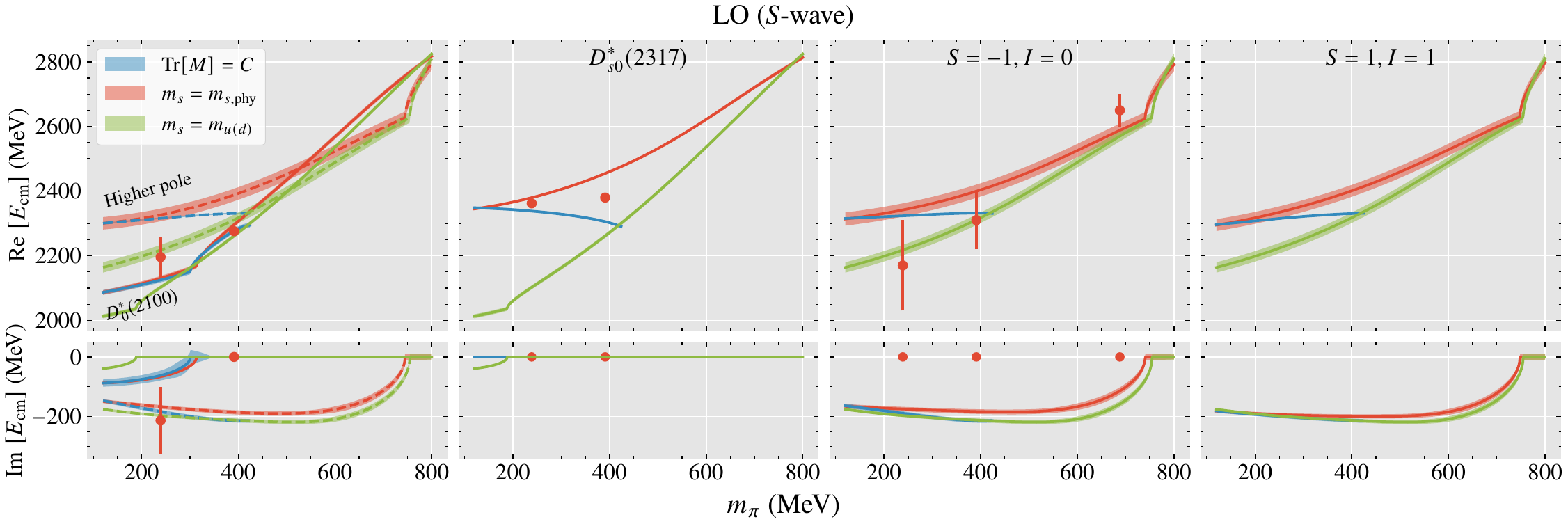}
	\vspace{4pt}
	\noindent\rule{0.9\textwidth}{0.4pt}
	\vspace{4pt}
	\includegraphics[width=0.9\textwidth]{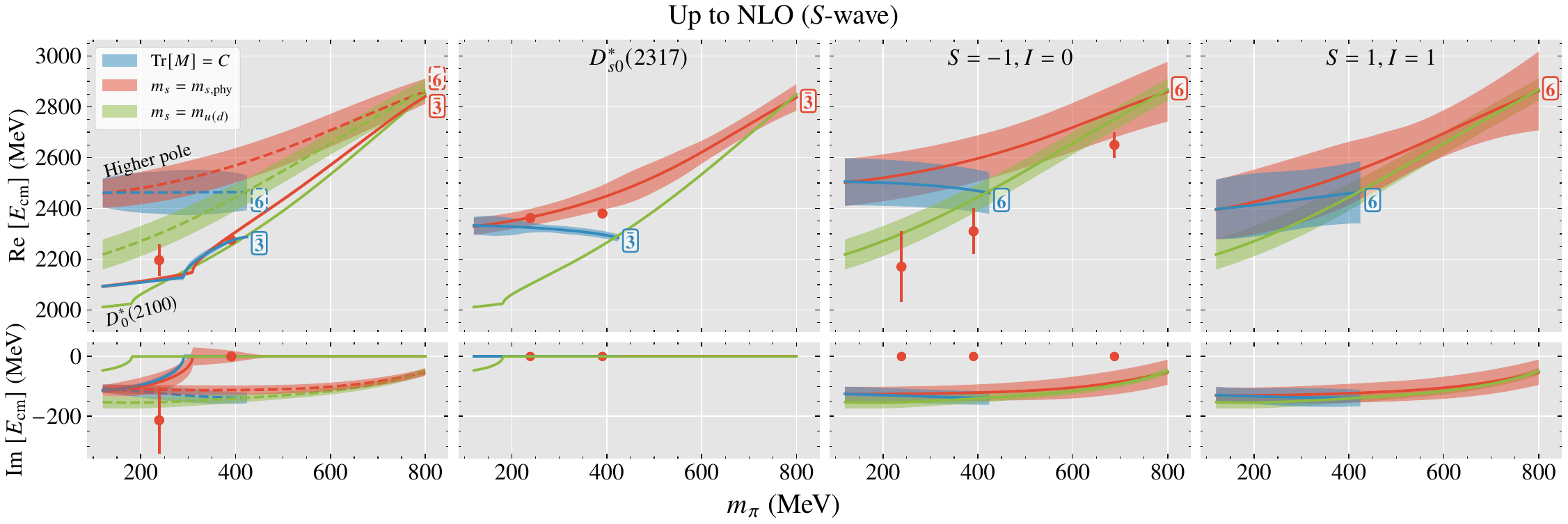}
	\caption{\label{fig:traj-pole-SI}Pole trajectories in different sectors $(S,I)$ by varying the $m_\pi$ at LO and up to NLO. The charm quark mass is fixed to its physical value.}
\end{figure*}

\end{widetext}
\end{document}